\documentclass[aps,prc,superscriptaddress,showpacs,amsfonts,twocolumn,floatfix]{revtex4-1}
\usepackage{color}
\usepackage{amsmath,amssymb}
\usepackage[makeroom]{cancel}
\usepackage{epsfig}
\usepackage{lineno}
\usepackage{soul}

\newcommand{\bl}[1]{{\color{blue} #1}}

\begin{document}
%\setpagewiselinenumbers
%\modulolinenumbers[2]
%\linenumbers

% Use the \preprint command to place your local institutional report
% number in the upper righthand corner of the title page in preprint mode.
% Multiple \preprint commands are allowed.
% Use the 'preprintnumbers' class option to override journal defaults
% to display numbers if necessary
%\preprint{}

%Title of paper
\title{Photoproduction of $\eta$~mesons off the proton for $1.2 < E_\gamma < 4.7$~GeV 
using CLAS at Jefferson Laboratory}

\newcommand{\comment}[1]{{\color{blue} #1}}

%\newcommand*{\FSU}{Florida State University, Tallahassee, Florida 32306, USA}
%\newcommand*{\FSUindex}{12}
%\affiliation{\FSU}
%\newcommand*{\VIRGINIA}{University of Virginia, Charlottesville, Virginia 22901, USA}
%\newcommand*{\VIRGINIAindex}{38}
%\affiliation{\VIRGINIA}

%\newcommand*{\UM}{University of Michigan, Ann Arbor, MI 48109}
%\newcommand*{\KAERI}{Korea Atomic Energy Research Institute, Gyeongju-si, 38180, South Korea}

%%%%%%%%%%%%%%% Latex Macros for institute addresses  %%%%%%%%%%%%%%%%%%%%%%%%%
\newcommand*{\ANL}{Argonne National Laboratory, Argonne, Illinois 60439, USA}
\newcommand*{\ANLindex}{1}
\affiliation{\ANL}
\newcommand*{\ASU}{Arizona State University, Tempe, Arizona 85287-1504, USA}
\affiliation{\ASU}
\newcommand*{\CSUDH}{California State University, Dominguez Hills, Carson, California 90747, USA}
\newcommand*{\CSUDHindex}{2}
\affiliation{\CSUDH}
\newcommand*{\CANISIUS}{Canisius College, Buffalo, New York 14208, USA}
\newcommand*{\CANISIUSindex}{3}
\affiliation{\CANISIUS}
\newcommand*{\CMU}{Carnegie Mellon University, Pittsburgh, Pennsylvania 15213, USA}
\newcommand*{\CMUindex}{4}
\affiliation{\CMU}
\newcommand*{\CUA}{Catholic University of America, Washington, DC 20064, USA}
\newcommand*{\CUAindex}{5}
\affiliation{\CUA}
\newcommand*{\SACLAY}{IRFU, CEA, Universit\'{e} Paris-Saclay, F-91191 Gif-sur-Yvette, France}
\newcommand*{\SACLAYindex}{6}
\affiliation{\SACLAY}
\newcommand*{\CNU}{Christopher Newport University, Newport News, Virginia 23606, USA}
\newcommand*{\CNUindex}{7}
\affiliation{\CNU}
\newcommand*{\UCONN}{University of Connecticut, Storrs, Connecticut 06269, USA}
\newcommand*{\UCONNindex}{8}
\affiliation{\UCONN}
\newcommand*{\DUKE}{Duke University, Durham, North Carolina 27708-0305, USA}
\newcommand*{\DUKEindex}{9}
\affiliation{\DUKE}
\newcommand*{\DUQUESNE}{Duquesne University, Pittsburgh, Pennsylvania 15282, USA}
\newcommand*{\DUQUESNEindex}{10}
\affiliation{\DUQUESNE}
\newcommand*{\FU}{Fairfield University, Fairfield, Connecticut 06824, USA}
\newcommand*{\FUindex}{11}
\affiliation{\FU}
\newcommand*{\FERRARAU}{Universit\`a di Ferrara, 44121 Ferrara, Italy}
\newcommand*{\FERRARAUindex}{12}
\affiliation{\FERRARAU}
\newcommand*{\FIU}{Florida International University, Miami, Florida 33199, USA}
\newcommand*{\FIUindex}{13}
\affiliation{\FIU}
\newcommand*{\FSU}{Florida State University, Tallahassee, Florida 32306, USA}
\newcommand*{\FSUindex}{14}
\affiliation{\FSU}
\newcommand*{\GWUI}{The George Washington University, Washington, DC 20052, USA}
\newcommand*{\GWUIindex}{15}
\affiliation{\GWUI}
\newcommand*{\ISU}{Idaho State University, Pocatello, Idaho 83209, USA}
\newcommand*{\ISUindex}{16}
\affiliation{\ISU}
\newcommand*{\INFNFE}{INFN, Sezione di Ferrara, 44100 Ferrara, Italy}
\newcommand*{\INFNFEindex}{17}
\affiliation{\INFNFE}
\newcommand*{\INFNFR}{INFN, Laboratori Nazionali di Frascati, 00044 Frascati, Italy}
\newcommand*{\INFNFRindex}{18}
\affiliation{\INFNFR}
\newcommand*{\INFNGE}{INFN, Sezione di Genova, 16146 Genova, Italy}
\newcommand*{\INFNGEindex}{19}
\affiliation{\INFNGE}
\newcommand*{\INFNRO}{INFN, Sezione di Roma Tor Vergata, 00133 Rome, Italy}
\newcommand*{\INFNROindex}{20}
\affiliation{\INFNRO}
\newcommand*{\INFNTUR}{INFN, Sezione di Torino, 10125 Torino, Italy}
\newcommand*{\INFNTURindex}{21}
\affiliation{\INFNTUR}
\newcommand*{\INFNPAV}{INFN, Sezione di Pavia, 27100 Pavia, Italy}
\newcommand*{\INFNPAVindex}{22}
\affiliation{\INFNPAV}
\newcommand*{\ORSAY}{Universit\'e Paris-Saclay, CNRS/IN2P3, IJCLab, 91405 Orsay, France}
\newcommand*{\ORSAYindex}{23}
\affiliation{\ORSAY}
\newcommand*{\Juelich}{Institute f\"ur Kernphysik, 52425 J\"ulich, Germany}
\newcommand*{\Juelichindex}{24}
\affiliation{\Juelich}
\newcommand*{\JMU}{James Madison University, Harrisonburg, Virginia 22807, USA}
\newcommand*{\JMUindex}{25}
\affiliation{\JMU}
\newcommand*{\KNU}{Kyungpook National University, Daegu 41566, Republic of Korea}
\newcommand*{\KNUindex}{26}
\affiliation{\KNU}
\newcommand*{\LAMAR}{Lamar University, 4400 MLK Blvd, P.O. Box 10009, Beaumont, Texas 77710, USA}
\newcommand*{\LAMARindex}{27}
\affiliation{\LAMAR}
\newcommand*{\MIT}{Massachusetts Institute of Technology, Cambridge, Massachusetts 02139-4307, USA}
\newcommand*{\MITindex}{28}
\affiliation{\MIT}
\newcommand*{\MISS}{Mississippi State University, Mississippi State, Mississippi 39762-5167, USA}
\newcommand*{\MISSindex}{29}
\affiliation{\MISS}
\newcommand*{\ITEP}{National Research Centre Kurchatov Institute - ITEP, Moscow, 117259, Russia}
\newcommand*{\ITEPindex}{30}
\affiliation{\ITEP}
\newcommand*{\UNH}{University of New Hampshire, Durham, New Hampshire 03824-3568, USA}
\newcommand*{\UNHindex}{31}
\affiliation{\UNH}
\newcommand*{\NSU}{Norfolk State University, Norfolk, Virginia 23504, USA}
\newcommand*{\NSUindex}{32}
\affiliation{\NSU}
\newcommand*{\OHIOU}{Ohio University, Athens, Ohio 45701, USA}
\newcommand*{\OHIOUindex}{33}
\affiliation{\OHIOU}
\newcommand*{\ODU}{Old Dominion University, Norfolk, Virginia 23529, USA}
\newcommand*{\ODUindex}{34}
\affiliation{\ODU}
\newcommand*{\RPI}{Rensselaer Polytechnic Institute, Troy, New York 12180-3590, USA}
\newcommand*{\RPIindex}{35}
\affiliation{\RPI}
\newcommand*{\URICH}{University of Richmond, Richmond, Virginia 23173, USA}
\newcommand*{\URICHindex}{36}
\affiliation{\URICH}
\newcommand*{\ROMAII}{Universit\`a di Roma Tor Vergata, 00133 Rome, Italy}
\newcommand*{\ROMAIIindex}{37}
\affiliation{\ROMAII}
\newcommand*{\MSU}{Skobeltsyn Institute of Nuclear Physics, Lomonosov Moscow State University, 119234 Moscow, Russia}
\newcommand*{\MSUindex}{38}
\affiliation{\MSU}
\newcommand*{\SCAROLINA}{University of South Carolina, Columbia, South Carolina 29208, USA}
\newcommand*{\SCAROLINAindex}{39}
\affiliation{\SCAROLINA}
\newcommand*{\TEMPLE}{Temple University,  Philadelphia, Pennsylvania 19122, USA}
\newcommand*{\TEMPLEindex}{40}
\affiliation{\TEMPLE}
\newcommand*{\JLAB}{Thomas Jefferson National Accelerator Facility, Newport News, Virginia 23606, USA}
\newcommand*{\JLABindex}{41}
\affiliation{\JLAB}
\newcommand*{\UTFSM}{Universidad T\'{e}cnica Federico Santa Mar\'{i}a, Casilla 110-V Valpara\'{i}so, Chile}
\newcommand*{\UTFSMindex}{42}
\affiliation{\UTFSM}
\newcommand*{\INSUBRIA}{Universit\`a degli Studi dell'Insubria, 22100 Como, Italy}
\newcommand*{\INSUBRIAindex}{43}
\affiliation{\INSUBRIA}
\newcommand*{\BRESCIA}{Universit\`{a} degli Studi di Brescia, 25123 Brescia, Italy}
\newcommand*{\BRESCIAindex}{44}
\affiliation{\BRESCIA}
\newcommand*{\GLASGOW}{University of Glasgow, Glasgow G12 8QQ, United Kingdom}
\newcommand*{\GLASGOWindex}{45}
\affiliation{\GLASGOW}
\newcommand*{\YORK}{University of York, York YO10 5DD, United Kingdom}
\newcommand*{\YORKindex}{46}
\affiliation{\YORK}
\newcommand*{\VT}{Virginia Tech, Blacksburg, Virginia 24061-0435, USA}
\newcommand*{\VTindex}{47}
\affiliation{\VT}
\newcommand*{\VIRGINIA}{University of Virginia, Charlottesville, Virginia 22901, USA}
\newcommand*{\VIRGINIAindex}{48}
\affiliation{\VIRGINIA}
\newcommand*{\WM}{College of William and Mary, Williamsburg, Virginia 23187-8795, USA}
\newcommand*{\WMindex}{49}
\affiliation{\WM}
\newcommand*{\YEREVAN}{Yerevan Physics Institute, 375036 Yerevan, Armenia}
\newcommand*{\YEREVANindex}{50}
\affiliation{\YEREVAN} 

\newcommand*{\NOWMISS}{Mississippi State University, Mississippi State, Mississippi 39762-5167, USA}
\newcommand*{\NOWOHIOU}{Ohio University, Athens, Ohio 45701, USA}
\newcommand*{\NOWISU}{Idaho State University, Pocatello, Idaho 83209, USA}
\newcommand*{\NOWBRESCIA}{Universit\`{a} degli Studi di Brescia, 25123 Brescia, Italy}
\newcommand*{\NOWJLAB}{Thomas Jefferson National Accelerator Facility, Newport News, Virginia 23606, USA}
\newcommand*{\NOWGWUI}{The George Washington University, Washington, DC 20052, USA}
%%%%%%%%%%%%%%% END OF Latex Macros for institute addresses %%%%%%%%%%%%%%%%%%%%%%%%% 

% repeat the \author .. \affiliation  etc. as needed
% \email, \thanks, \homepage, \altaffiliation all apply to the current
% author. Explanatory text should go in the []'s, actual e-mail
% address or url should go in the {}'s for \email and \homepage.
% Please use the appropriate macro foreach each type of information

% \affiliation command applies to all authors since the last
% \affiliation command. The \affiliation command should follow the
% other information
% \affiliation can be followed by \email, \homepage, \thanks as well.
\author{T.~Hu} \affiliation{\FSU}
\author{Z.~Akbar} \altaffiliation[Present address: ]{\VIRGINIA} \affiliation{\FSU}
%\author{P.~Roy} \altaffiliation[Present address: ]{\UM}\affiliation{\FSU}
%\author{S.~Park} \altaffiliation[Present address: ]{\KAERI}\affiliation{\FSU}
\author{V.~Crede} \altaffiliation[Corresponding author: crede@fsu.edu]{} \affiliation{\FSU}
%\author{A.~V.~Anisovich} \affiliation{\BONN} \affiliation{\NRC}
%\author{I.~Denisenko} \affiliation{\BONN} \affiliation{\JINR}
%\author{E.~Klempt} \affiliation{\BONN} \affiliation{\JLAB}
%\author{V.~A.~Nikonov} \affiliation{\BONN} \affiliation{\NRC}
%\author{A.~V.~Sarantsev} \affiliation{\BONN} \affiliation{\NRC}

\author {K.P.~Adhikari} 
\altaffiliation[Present address: ]{\NOWMISS}
\affiliation{\ODU}
\author {S.~Adhikari} 
\affiliation{\FIU}
\author {M.J.~Amaryan} 
\affiliation{\ODU}
\author {G.~Angelini} 
\affiliation{\GWUI}
\author {G.~Asryan} 
\affiliation{\YEREVAN}
\author {H.~Atac} 
\affiliation{\TEMPLE}
\author {C.~Ayerbe Gayoso} 
\affiliation{\WM}
% ...

%\email[]{Your e-mail address}
%\homepage[]{Your web page}
%\thanks{}
%\altaffiliation{}
%\affiliation{}

%Collaboration name if desired (requires use of superscriptaddress
%option in \documentclass). \noaffiliation is required (may also be
%used with the \author command).
%\collaboration can be followed by \email, \homepage, \thanks as well.
\collaboration{The CLAS Collaboration}
\noaffiliation

\date{Received: \today / Revised version:}

\begin{abstract}
Photoproduction cross sections are reported for the reaction $\gamma p\to p\eta$ using energy-tagged
photons and the CLAS spectrometer at Jefferson Laboratory. The $\eta$~mesons are detected in their dominant
charged decay mode, $\eta\to \pi^+\pi^-\pi^0$, and results on differential cross sections are presented 
for incident photon energies between 1.2~and 4.7~GeV. These new $\eta$~photoproduction data are consistent
with earlier CLAS results but extend the energy range beyond the nucleon resonance region into the Regge
regime. The normalized angular distributions are also compared with the experimental results from several 
other experiments, and with predictions of $\eta$-MAID\,2018 and the latest solution of the Bonn-Gatchina 
coupled-channel analysis. Differential cross sections $d\sigma/dt$ are presented for incident photon
energies $E_\gamma > 2.9$~GeV ($W > 2.5$~GeV), and compared with predictions which are based on Regge 
trajectories exchange in the $t$-channel (Regge models). The data confirm the expected dominance 
of $\rho$, $\omega$ vector-meson exchange in an analysis by the Joint Physics Analysis Center.
\end{abstract}

% insert suggested PACS numbers in braces on next line
%\pacs{13.60.Le Meson production, 
%      13.60.-r Photon and charged-lepton interactions with hadrons,  
%      14.20.Gk Baryon resonances,
%      25.20.Lj Photoproduction reactions}
% insert suggested keywords - APS authors don't need to do this
%\keywords{}

%\maketitle must follow title, authors, abstract, \pacs, and \keywords
\maketitle

% body of paper here - Use proper section commands
% References should be done using the \cite, \ref, and \label commands
\section{\label{Section:Introduction}Introduction}
% Put \label in argument of \section for cross-referencing
%\section{\label{}}

The photoproduction of pseudoscalar mesons on the nucleon has remained of interest in recent years for the
study of meson production in hadronic reactions across a wide range of energies. At low energies using incident 
photon energies below 3.0~GeV, information about the nucleon excitation spectrum can be extracted, whereas at 
higher energies above $E_\gamma\approx 4$~GeV, details of the residual hadron interactions due to the $t$-channel 
exchange of massive quasi-particles known as Reggeons can be studied~\cite{Irving:1977ea}. These two regimes are 
analytically connected, but the scarcity of cross section and polarization data for the energy range 3--6~GeV 
have thus far hindered our understanding of the transition from the baryon resonance regime to high-energy 
photoproduction. While each Reggeon exchange has a known energy behavior, the dependence on the momentum
exchange in the reaction is initially unknown. However, dispersion relations can be used to derive the 
$t$-dependence of the Regge amplitudes from the low-energy amplitude, which is usually described in terms of a 
finite number of partial waves. This technique of finite-energy sum rules (FESR) was recently applied to $\pi^0$ 
and $\eta$~photoproduction~\cite{Mathieu:2018mjw,Nys:2016vjz}. In these reactions, the resonance and the Regge 
regions can be effectively separated. Low-energy amplitudes, which directly contain the resonance dynamics, 
should then smoothly connect with the high-energy region~\cite{Tiator:2018heh}. Alternatively, FESR can be 
derived to constrain the low-energy amplitudes by the $t$-channel Reggeon exchanges~\cite{Mathieu:2015eia} 
and ultimately, to extract nucleon resonance parameters. Both approaches were recently explored for $\pi^0$ 
photoproduction in Ref.~\cite{Mathieu:2015gxa}.

In the nucleon resonance region, abundant data on $\eta$~photoproduction on the proton are available from the
reaction threshold at $W_{\rm \, thres.}\approx 1.49$~GeV up to the fourth resonance region just below $W\approx
2$~GeV. The data situation has even improved in recent years, particularly for (double-)polarization observables
with the availability of longitudinally and transversely polarized targets at several photoproduction facilities 
around the world, e.g., Jefferson Laboratory~\cite{Mecking:2003zu} in Newport News, USA, ELSA~\cite{Hillert:2006yb}
in Bonn, Germany, and MAMI~\cite{Mecking:2006yh} in Mainz, Germany. Data using high-intensity photon beams with 
excellent linear polarization are also available from the GRAAL facility~\cite{Bartalini:2005wx} in Grenoble, 
France, and from LEPS~\cite{Muramatsu:2013tla} at SPring-8 in Hyogo, Japan. The photo-induced production of 
$\eta$~mesons is a selective probe for the study of nucleon excitations. Although photons incident on protons 
couple to both isospin $I=0,1$ initial states, the $\eta$~meson in the final state serves as an isospin filter 
for baryon excitations since isospin $I=3/2$~states ($\Delta$~resonances) are prohibited from decaying into 
$N\eta$~final states.

Near the production threshold, the dominance of the two nucleon resonances $N(1535)\,1/2^-$ and
$N(1650)\,1/2^-$ in $\eta$~photoproduction is undisputed~\cite{Krusche:1995nv,Krusche:1997jj}. Small 
contributions have also been observed in $(\gamma,\eta)$ from the $N(1520)\,3/2^-$~state, which itself couples
strongly to the $N\eta$~decay mode. The state was identified mainly from the $S_{11}$\,-\,$D_{13}$ interference
term in the description of the photon-beam asymmetry~\cite{Ajaka:1998zi,Elsner:2007hm,Bartalini:2007fg,
Collins:2017sgu} indicating the importance of polarization observables. Also available are results from MAMI 
for the transverse target asymmetry~$T$, and the beam-target asymmetry~$F$~\cite{Akondi:2014ttg}. The helicity
asymmetry~$E$ was reported by the CLAS Collaboration at Jefferson Lab~\cite{Senderovich:2015lek} and the A2
Collaboration at MAMI~\cite{Witthauer:2017wdb}. More recently, results on the target asymmetry~$T$ and the 
double-polarization observables $E$, $G$ (longitudinal target polarization) as well as $P$, $H$ (transverse 
target polarization) in the photoproduction of $\eta$~mesons off protons were reported by the CBELSA/TAPS 
Collaboration at ELSA~\cite{Muller:2019qxg}.

In their bi-annual editions, the listing of nucleon resonances by the 
Particle Data Group (PDG) in the {\it Review of Particle Physics}~\cite{Tanabashi:2018oca} has undergone 
significant upgrades based on the recent photoproduction data from the above facilities with almost no 
$N^\ast$~resonance left untouched since 2010. Several new nucleon states have been added, some of which 
show strong couplings to $N\eta$. Above 1700~MeV in overall center-of-mass energy, a third $1/2^-$~state, 
$N(1895)\,1/2^-$, is now listed as a new resonance with a four-star rating indicating its existence is 
certain in both its overall status and its $N\eta$~decay mode. In the $1/2^+$~wave, a large contribution 
in $(\gamma,\eta)$ is observed from the $N(1710)\,1/2^+$~resonance, the status of which has been upgraded 
to three stars in its $N\eta$ decay mode. In the fourth resonance region and above, discrepancies occur in 
various amplitude analyses. Such ambiguities are not surprising in light of the remaining {\it incompleteness} 
of the $\eta$~photoproduction database. The experimental status of $\eta$~photoproduction from nucleons and 
nuclei, as well as phenomenological progress was recently reviewed in Ref.~\cite{Krusche:2014ava}.

\begin{figure}[t]
 \includegraphics[width=0.40\textwidth]{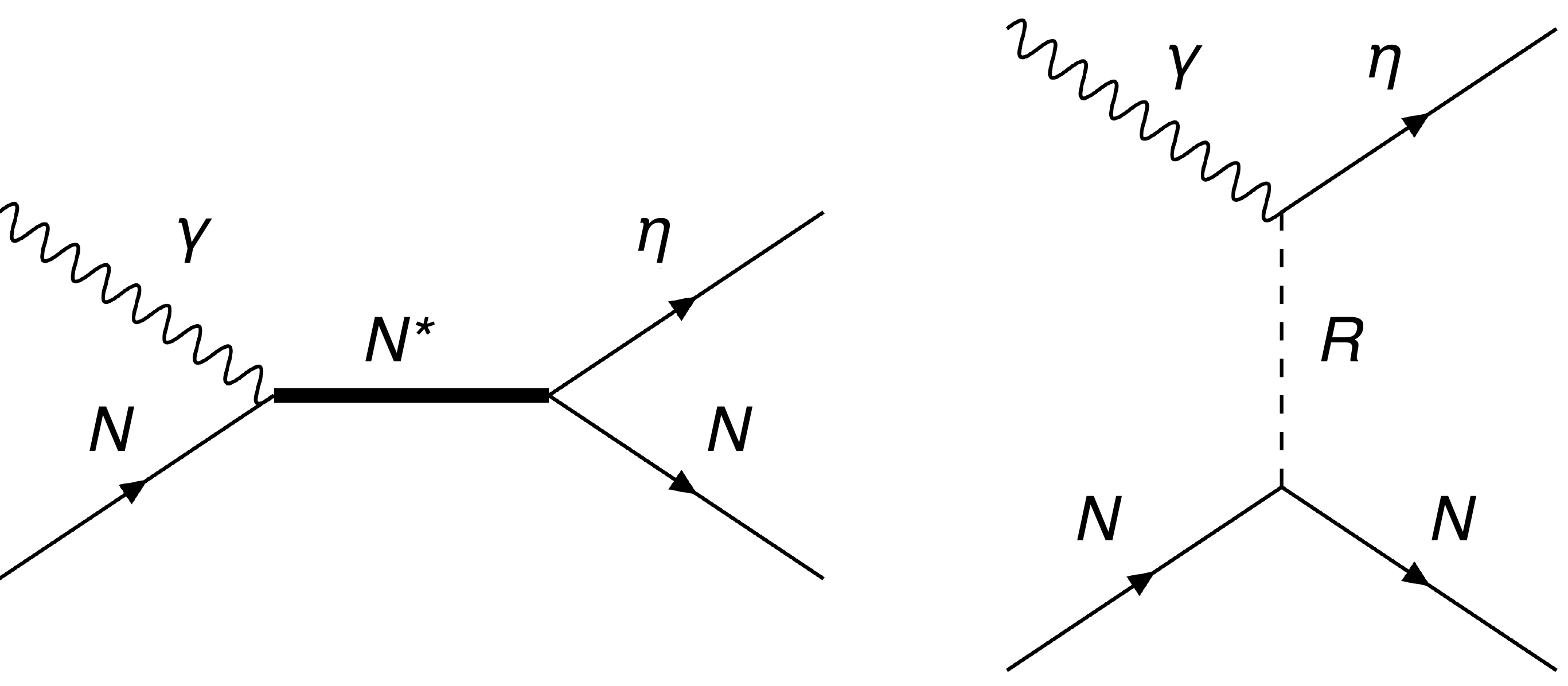}
 \caption{\label{Figure:feynman} Dominant contributions to $\eta$~photoproduction off the nucleon: $s$-channel 
   intermediate nucleon resonance excitation (left) and $t$-channel exchange of Reggeons (right).}
\end{figure}

The theoretical description of high-energy photoproduction provides constraints on the amplitudes utilized in
low-energy meson photoproduction to extract the spectrum of excited baryons~\cite{Mathieu:2015gxa}. Moreover,
understanding the meson photoproduction mechanism at high energies is a crucial component of a broader program
to search for gluonic excitations in the meson spectrum, which is the primary goal of the GlueX experiment in 
Hall~D at Jefferson Lab~\cite{AlGhoul:2017nbp,Adhikari:2020cvz}. 

In a brief summary, $\eta$~photoproduction off the nucleon is dominantly described by $s$-channel intermediate 
nucleon resonance excitation (left side of Fig.~\ref{Figure:feynman}) close to the production threshold with an 
admixture of $t$-channel exchange of Reggeons (right side of Fig.~\ref{Figure:feynman}) as the incident photon 
energy increases, whereas the $t$-channel Reggeon exchange dominates the production of $\eta$~mesons at higher 
energies above 7~GeV.

In this paper, differential cross sections are presented for the reaction $\gamma p\to p\eta$ from CLAS at 
Jefferson Lab, where the $\eta$ was identified through the detection of its decay products $\pi^+\pi^-\pi^0$. 
The new data reported here cover an incident photon energy range $E_\gamma$ from 1.2~GeV up to 4.7~GeV.

This paper has the following structure. A summary of previous measurements in $\eta$~photoproduction is
presented in Sec.~\ref{Section:PreviousResults}. Section~\ref{Section:ExperimentalSetup} gives an introduction 
to the CLAS-g12 experimental setup. The data reconstruction and event selection are discussed in
Sec.~\ref{Section:Selection} and the extraction of the cross sections is described in 
Sec.~\ref{Section:Extraction}. Finally, the experimental results and a discussion of the physics including 
possible nucleon resonance contributions are presented in Secs.~\ref{Section:Results} and~\ref{Section:PWA}, 
respectively.

% If in two-column mode, this environment will change to single-column
% format so that long equations can be displayed. Use
% sparingly.
%\begin{widetext}
% put long equation here
%\end{widetext}

\section{\label{Section:PreviousResults}Previous Measurements}
Cross sections for the reaction $\gamma p\to p\eta$ were measured at many different laboratories over a wide 
kinematic range and in various $\eta$~decay modes using either tagged-photon beams produced in Compton scattering 
of laser photons off electrons in the accelerator~\cite{Renard:2000iv,Bartalini:2007fg,Sumihama:2009gf} or via 
the bremsstrahlung technique~\cite{Crede:2003ax,Bartholomy:2007zz,Williams:2009yj,Crede:2009zzb,McNicoll:2010qk,
Kashevarov:2017kqb}. A summary of the experimental data on $\eta$~photoproduction cross sections from the nucleon 
is given in Table~\ref{Table:Measurements}. The current status of single-$\eta$~meson production using photon 
beams is reviewed in Ref.~\cite{Ireland:2019uwn}, and in particular the information that can be obtained on the 
spectrum of light, non-strange baryons.

A whole ``industry'' of photoproduction experiments recorded data for several meson-production channels in 
the 60s and 70s. Results were mostly published at higher energies and only a few data points bridge the
gap down to the resonance region below $E_\gamma\approx 3$~GeV. Particularly interesting for the discussion
of cross sections is the normalization technique of these older data since tagged-photon beams were not 
available at these facilities. 

\subsection{Summary of older photoproduction experiments (1960s and 1970s)}
At the 5~GeV electron synchrotron NINA at the Daresbury Laboratory, a linearly polarized bremsstrahlung 
beam was used to extract differential cross sections for the reaction $\gamma p\to p\eta$ at incident 
photon energies of 2.5 GeV and 3.0 GeV, and for various $t$-values between $-0.2$~GeV$^2$ and 
$-1.2$~GeV$^2$~\cite{Bussey:1976si}. The incident photon intensity as a function of energy was derived 
from a quantameter, together with the shape of the spectrum as measured with a pair spectrometer. At the 
Deutsches Elektronen-Synchrotron (DESY), a bremsstrahlung beam was produced on a tungsten target and the 
flux was measured with a gas-filled quantameter. Cross section results for $\eta$~photoproduction were 
reported at mean photon energies of 4 and 6~GeV in the momentum transfer range between zero and 
1.4~GeV$^2$~\cite{Braunschweig:1970jb}. 

A bremsstrahlung beam from a tungsten target was used at the Cambridge Electron Accelerator (CEA) at the 
Massachusetts Institute of Technology (MIT). The beam was monitored with a quantameter that was calibrated 
against a Faraday cup and whose output was measured with a current integrator~\cite{Elings:1967af}. Results 
for $\eta$~photoproduction at 4~GeV were published in Ref.~\cite{Bellenger:1968zz}. Finally, cross section
measurements were also performed at the 10~GeV synchrotron at the Laboratory of Nuclear Studies at Cornell
University. Several data points were published for incident photon energies of 4~and 8~GeV and momentum 
tranfers $-t$ between 0.3 and 0.8~GeV$^2$~\cite{Dewire:1972kk}.

Measurements at higher incident photon energies in the range 6.0--16.0~GeV were performed at the Stanford 
Linear Accelerator Center (SLAC) using a bremsstrahlung beam~\cite{Anderson:1968wy}. The beam was monitored 
by detecting Cherenkov light of $e^+e^-$~pairs from a converter in the beam. The Cherenkov monitor was 
calibrated against a precision calorimeter~\cite{Boyarski:1967sp}. In Ref.~\cite{Boyarski:1969iy}, the overall 
uncertainty in normalization was estimated at 10\,\%; other references give even smaller uncertainties, see e.g. 
Ref.~\cite{Boyarski:1967sp}. The SLAC high-power quantameter was used for the measurement of the incident photon 
flux and is described in Ref.~\cite{Anderson:1968tj}.

\subsection{Experiments using Compton backscattering}
The GRenoble Anneau Accelerateur Laser (GRAAL) experiment measured the differential $\eta$~photoproduction 
cross sections from threshold up to 1100~MeV~\cite{Renard:2000iv} and up to 1500~MeV~\cite{Bartalini:2007fg} 
in incident photon laboratory energy and for cos\,$\theta_{\rm \,c.m.} < 0.85$ of the $\eta$~meson in the 
overall center-of-mass (c.m.) frame. The facility was located at the European Synchrotron Radiation Facility 
(ESRF) in Grenoble, France. For a detailed description of the facility, see Ref.~\cite{Bartalini:2005wx}. The 
tagged and polarized $\gamma$-ray beam was produced by Compton scattering of laser photons off the 6~GeV electrons 
circulating in the storage ring. The photon energy was provided by an internal tagging system consisting of silicon
microstrips for the detection of the scattered electron and a set of plastic scintillators for time-of-flight (TOF) 
measurements~\cite{Bartalini:2007fg}. A thin monitor was used to measure the beam flux (typically $10^6~\gamma$/s).
The monitor efficiency of $(2.68\pm 0.03)\,\%$ was estimated by comparing with the response of a lead/scintillating
fiber calorimeter at a low rate. 

At the SPring-8/LEPS facility, the photon beam was produced by backward-Compton scattering of laser photons off 
electrons with an energy of 8~GeV. Data were accumulated with $1.0\times 10^{12}$~photons at the target and cross 
section results on the reaction $\gamma p\to p\eta$ were extracted for the incident photon energy range 
$E_\gamma\in [\,1.6,\,2.4\,]$~GeV in the backward direction 
(cos\,$\theta_{\rm \,c.m.} < -0.6$)~\cite{Sumihama:2009gf}. 

\begin{table}[b]
\begin{tabular}{c|c|c|c}
  Reaction~\,&~$W$ [\,GeV\,]~&~$-t$ [\,GeV$^2$\,]~&~Reference\\[0.5ex]\hline
  $\gamma p\to p\eta$ &\,1.49\,--\,1.96\,& $-$ & A2~\cite{McNicoll:2010qk,Kashevarov:2017kqb}\\[0.1ex]
  %$\gamma p\to p\eta$ 
  &\,1.55\,--\,2.80\,& $-$ & {\small CLAS~\cite{Dugger:2002ft,Williams:2009yj}}\\[0.1ex]
  %$\gamma p\to p\eta$ 
  &\,1.51\,--\,2.55\,& $-$ & {\small CB-ELSA~\cite{Crede:2003ax,Bartholomy:2007zz}}\\[0.1ex]
  %$\gamma p\to p\eta$ 
  &\,1.57\,--\,2.38\,& $-$ &\,{\small CBELSA/TAPS~\cite{Crede:2009zzb}}\\[0.1ex]
  %$\gamma p\to p\eta$ 
  &\,1.49\,--\,1.92\,& $-$ & {\small GRAAL~\cite{Bartalini:2007fg,Renard:2000iv}}\\[0.1ex]
  %$\gamma p\to p\eta$ 
  &\,1.97\,--\,2.32\,& $-$ & {\small LEPS~\cite{Sumihama:2009gf}}\\[0.1ex]
  %$\gamma p\to p\eta$ 
  &\,2.36 \& 2.55\,& 0.2\,--\,1.2 & {\small Daresbury~\cite{Bussey:1976si}}\\[0.1ex]
  %$\gamma p\to p\eta$ 
  &\,2.90 \& 3.48\,& 0.0\,--\,1.4 & {\small DESY~\cite{Braunschweig:1970jb}}\\[0.1ex]
  %$\gamma p\to p\eta$ 
  &\,2.90 & $< 1.0$ & {\small MIT~\cite{Bellenger:1968zz}}\\[0.1ex]
  %$\gamma p\to p\eta$ 
  &\,3.48\,--\,5.56\,& 0.2\,--\,0.9 & {\small SLAC~\cite{Anderson:1968wy}}\\[0.1ex]
  %$\gamma p\to p\eta$ 
  &\,2.90\,--\,3.99\,& 0.3\,--\,0.8 & {\small Cornell~\cite{Dewire:1972kk}}\\[0.5ex]\hline
  $\gamma n\to n\eta$ &\,1.49\,--\,1.88\,& $-$ & A2~\cite{Werthmuller:2014thb}\\[0.1ex]
  %$\gamma n\to n\eta$ 
  &\,1.50\,--\,2.18\,& $-$ &\,{\small CBELSA/TAPS~\cite{Jaegle:2011sw}}\\[0.1ex]
  %$\gamma n\to n\eta$ 
  &\,1.59\,--\,2.07\,& $-$ &\,{\small CBELSA/TAPS~\cite{Witthauer:2017pcy}}
\end{tabular}
\caption{\label{Table:Measurements}Summary of experimental data on cross sections for
  $\eta$~photoproduction off the nucleon.}
\end{table}

\begin{figure*}[t]
 \includegraphics[width=0.99\textwidth]{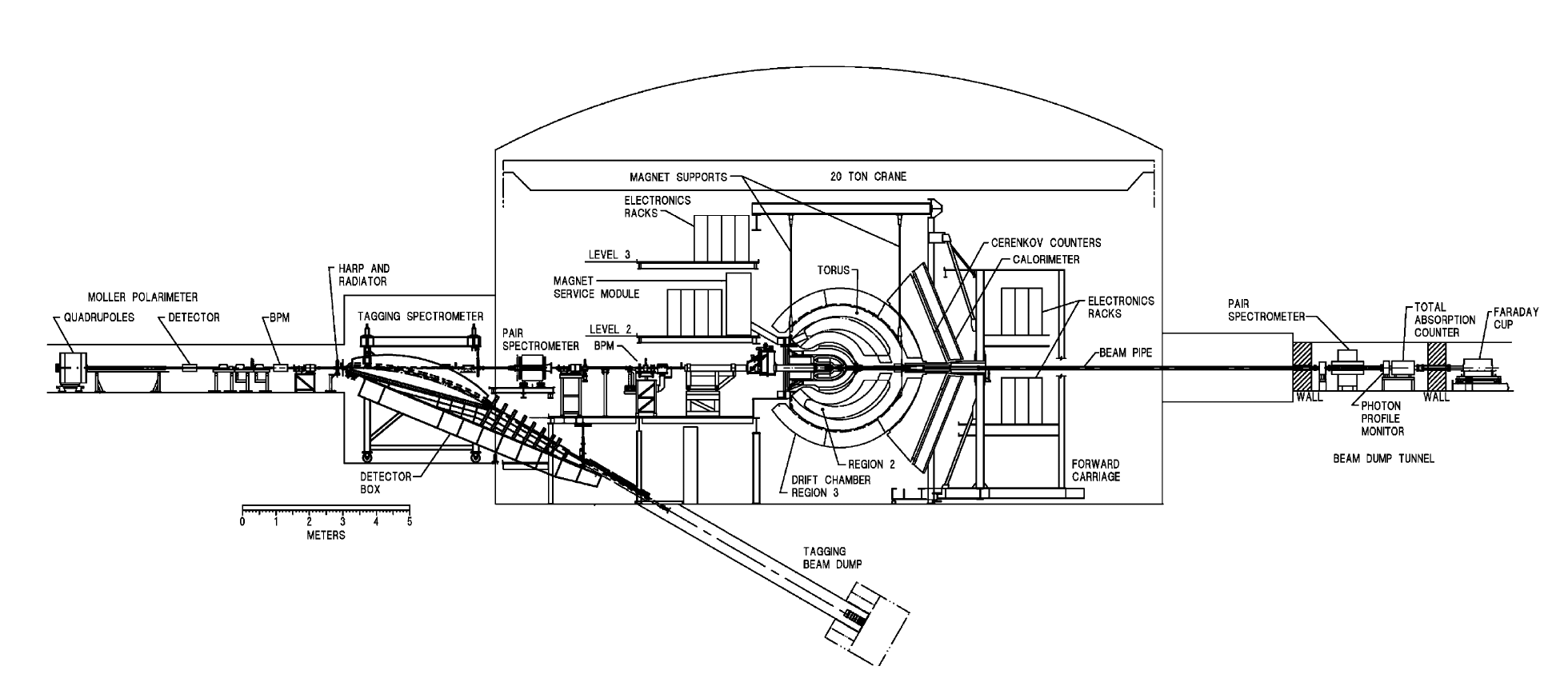} 
 \caption{\label{Figure:CLAS} Side view of the CLAS detector in Hall B at Jefferson Lab including the
   photon tagging facility upstream of CLAS. Reproduced figure with permission from
   Ref.~\cite{Mecking:2003zu}. Copyright 2003 by Elsevier.}
\end{figure*}

\subsection{Experiments using bremsstrahlung photons}
At the ELectron Stretcher Accelerator (ELSA)~\cite{Hillert:2006yb}, two very different experimental 
setups extracted cross section data for the photo-produced $p\eta$~final state. In 2001, the CB-ELSA 
detector recorded data and $\eta$~photoproduction was studied in the neutral decays of the $\eta$~meson 
into $\gamma\gamma$ and $\pi^0\pi^0\pi^0$~\cite{Crede:2003ax,Bartholomy:2007zz}. The original 
experiment consisted of the CsI(Tl)-based Crystal Barrel (CB) calorimeter covering 97.8\,\% of
the $4\pi$~solid angle~\cite{Aker:1992ny}. For the 2000/2001 data taking, electrons were extracted 
in two separate experiments at energies of 1.4 and 3.2~GeV, covering tagged-photon energies from 
0.3~up to about 3.0~GeV, with a typical intensity of 1--3\,$\times$\,$10^6$ tagged photons/s. The 
experimental setup was later modified and in a series of measurements in 2002/2003, a combination 
of the CB calorimeter and the BaF$_2$~TAPS detector in the forward direction was used. Results 
of the CBELSA/TAPS setup on single-$\eta$ cross section measurements off the proton can be found 
in Ref.~\cite{Crede:2009zzb}. The data provide improved angular coverage in the forward and backward 
direction in the center-of-mass system.

At the upgraded Mainz Microtron (MAMI-C), an experimental setup using a combination of the NaI(Tl) 
Crystal Ball and BaF$_2$ TAPS multi-photon spectrometers recorded high-quality data on the reaction 
$\gamma p\to p\eta$ in the energy range from the production threshold at 707~MeV to 1.4~GeV
\cite{McNicoll:2010qk, Kashevarov:2017kqb}. The NaI(Tl)~crystals were arranged in two hemispheres
that covered 93\,\% of the $4\pi$~solid angle and the TAPS calorimeter subtended the full azimuthal
range for polar angles from $1^\circ$ to $20^\circ$. Since the TAPS calorimeter was installed 1.5~m
downstream of the Crystal Ball center, the resolution of TAPS in the polar angle~$\theta$ was better than
$1^\circ$. For an electron beam energy of 1508~MeV, a tagger channel in this experiment had a width of
about 2~MeV at $E_\gamma = 1402$~MeV and about 4~MeV at the $\eta$-photoproduction threshold of 
$E_\gamma = 707$~MeV.

At the Continuous Electron Beam Accelerator Facility (CEBAF) at Jefferson Laboratory (Jefferson 
Lab), the CEBAF Large Acceptance Spectrometer (CLAS) was optimized for charged-particle tracking. 
A detailed description of the spectrometer and its various detector components is given below and in 
Ref.~\cite{Mecking:2003zu}. The CLAS ``g1'' experiment accumulated data in 1998 (g1a) and in 1999 
(g1c) using electron beam energies of 2.49~and 2.45~GeV, respectively. These experiments used a 
single-prong trigger configuration. Results for the reaction $\gamma p\to p\eta$ were only published 
from the CLAS ``g1a'' experiment~\cite{Dugger:2002ft}. For the absolute normalization of the $\eta$
channel, the SAID-SM02 solution~\cite{SAID-SM02} was used. The normalization uncertainty for all 
incident photon energies below 2~GeV was estimated at~3\,\%~\cite{Dugger:2002ft}. 

The CLAS ``g11a'' experiment accumulated a high-statistics data sample in 2004 of about $20\times
10^9$~triggered events. An electron beam of energy $E_{\rm \,e^-} = 4.023$~GeV was used to generate 
tagged photons with energies between 0.81 and 3.81~GeV covering center-of-mass energies up to $\sqrt{s}
\approx 2.84$~GeV. Results on $\eta$~cross section measurements for $E_\gamma < 3.5$~GeV are
published in Ref.~\cite{Williams:2009yj}.

A review of the main photoproduction data sets prior to 2013 and a corresponding comparison 
of their coverage in energy and solid angle can be found in Ref.~\cite{Crede:2013sze}. 

\section{\label{Section:ExperimentalSetup}Experimental Setup}
The $\gamma p\to p\eta$ measurements discussed here were performed at Jefferson Lab from March to June
2008 using the CLAS spectrometer~\cite{Mecking:2003zu} in Hall~B. The experimental setup is shown in
Fig.~\ref{Figure:CLAS}. The incident tagged, bremsstrahlung photon beam was produced from a 60--65~nA
electron beam of energy $E_{\rm\,e^-}=5.715$~GeV delivered by the \text{CEBAF} accelerator. These measurements
were part of the CLAS-g12 experiment, which was a high-luminosity data-taking period. The tagging system
provided a circularly polarized, real-photon  beam with the highest available photon energies of any CLAS
experiment of up to $E_\gamma\approx 5.4$~GeV, corresponding to about 95\,\% of $E_{\rm\,e^-}$. The photons
impinged upon a 40-cm-long unpolarized liquid-hydrogen target, which was moved upstream by 90~cm from the
center of the CLAS spectro\-meter to enhance the acceptance of charged tracks in the forward direction. 
Various results from the CLAS-g12 experiment have been recently published and are discussed in
Refs.~\cite{Chandavar:2017lgs,Kunkel:2017src,Bono:2018ike,Goetz:2018ynn}. First cross section measurements
have been presented in short papers on the reaction
$\gamma p\to p\pi^0\to pe^+ e^-\,(\gamma)$~\cite{Kunkel:2017src} and on the reaction 
$\gamma p\to K^+ K^+\,(X)$~\cite{Goetz:2018ynn} in the search for excited $\Xi$~baryons.

A brief overview of the CLAS performance is given in the following section; a full description of 
the CLAS spectrometer can be found in Ref.~\cite{Mecking:2003zu}. The remaining sections describe 
at greater length those components of the experimental setup that differ from previous CLAS 
experiments or are particularly relevant for the cross section measurements.

\subsection{\label{Subsection:Overview}Overview} 
The charged tracks in the experiment were detected in the CLAS spectrometer, which provided coverage 
for charged particles in the polar-angle range $8^\circ < \theta_{\rm \,lab} < 135^\circ$. The three
momentum components of the particles were reconstructed from their tracks in the toroidal magnetic 
field of the spectrometer by a set of three drift-chamber packages~\cite{Mestayer:2000we}. Time-of-flight 
(TOF) information was available from plastic scintillators~\cite{Smith:1999ii} located about 5~m from 
the center of CLAS. The spectrometer provided a momentum and angle resolution of $\Delta p/p\approx 
1\,\%$ and $\Delta\theta\approx 1^\circ$\,--\,$2^\circ$, respectively. A set of plastic scintillation 
counters close to the target (referred to as the start counter) provided event start 
times~\cite{Sharabian:2005kq}. For this experiment, coincident signals from the photon tagger, start 
counter, and time-of-flight system constituted the event trigger that required a coincidence between 
a scattered-electron signal from the photon tagger and an energy-dependent number of charged tracks 
in CLAS (see Sec.~\ref{Subsection:Trigger} for details).

\subsection{\label{Subsection:TaggingSystem}The tagging system}
The bremsstrahlung beam was produced from a thin gold radiator and photons were tagged by detecting
energy-degraded electrons, which were deflected in the magnetic field of a single dipole magnet. The
CLAS tagging system used a hodoscope that contained two planar arrays of plastic scintillators
\cite{Sober:2000we}. The first layer of 384 partially overlapping small scintillators (E-counters) 
provided the photon energy accuracy of \mbox{$\approx 1\times 10^{-3}\,E_{\rm \,e^-}$}, while the 
second layer of 61 larger scintillators (T-counters) provided the timing resolution of about 160~ps 
necessary to form a coincidence with the corresponding charged particles that were produced in the 
nuclear interaction triggered by the tagged photon.

The arrangement of the E-counters is relevant for the discussion of the cross-section results presented 
here. The widths of the counters ranged from 6 to 18~mm to provide approximately constant
momentum intervals of $0.003\,E_{\rm \,e^-}$. Since each counter optically overlapped its adjacent
neighbors by one-third of their respective widths, a total of 767~separate incident photon energy bins
was available with an energy range of approximately $r=0.001\,E_{\rm \,e^-}$. Assuming equal acceptance 
along the length of each paddle, the element for the photon energy in the covariance matrix is given by:
\begin{eqnarray}
\sigma^2_{E_\gamma} = \frac{1}{r}~\int\limits^{r/2}_{-r/2}\,E^2\,dE \,=\, \frac{r^2}{12}\,.
\end{eqnarray}
The CLAS-g12 experiment recorded data at the highest possible CEBAF energies of $E_{\rm \,e^-} = 5.715$~GeV
and therefore, $\sigma_{E_\gamma} = 1.65$~MeV, which is about 80\,\% greater than the number for the CLAS-g11a 
experiment~\cite{Williams:2009yj}. Thus, a slightly broader binning in center-of-mass energy~$W$ was chosen 
for $W < 2.1$~GeV. In particular, very close to the low-energy end of the tagging range at about 21\,\% 
of $E_{\rm \,e^-}$, the width of the $W$~bins translates into the smallest bin width in incident photon 
energy for the entire analyzed energy range. This resolution effect, combined with observed small 
fluctuations in our extracted cross sections at the lower end of the tagging range, which are believed 
to originate from the measured incident photon flux, required adjusting the $W$~binning from 20~MeV to 
40~MeV for $W < 1.88$~GeV.

\subsection{\label{Subsection:PID}Particle identification}
Particle identification (PID) of charged final-state hadrons in this experiment was based on the 
combined information from the drift chamber and TOF systems. A value for $\beta$, defined as the 
ratio of the particle speed relative to the speed of light, could be measured in two different ways: 
\begin{enumerate}
\item An empirically measured value for each particle, $\beta_{\,m} = v/c = \Delta t/(c\,l)$, was 
  based on timing information from the time-of-flight and start counter systems (where $l$ denotes 
  the length of the track as determined from the drift-chamber track reconstruction), and
\item Independently, a value for each particle, $\beta_{\,c} = p/E$, could be determined from the 
  measured momentum using the CLAS drift chambers and the PDG mass~\cite{Tanabashi:2018oca} for the 
  particle.
\end{enumerate}
PID could then proceed by evaluating the distribution of $\Delta\beta=|\beta_{\,c}-\beta_{\,m}|$
values and defining proper event-by-event selection criteria.

\begin{figure}[t]
 \includegraphics[width=0.5\textwidth]{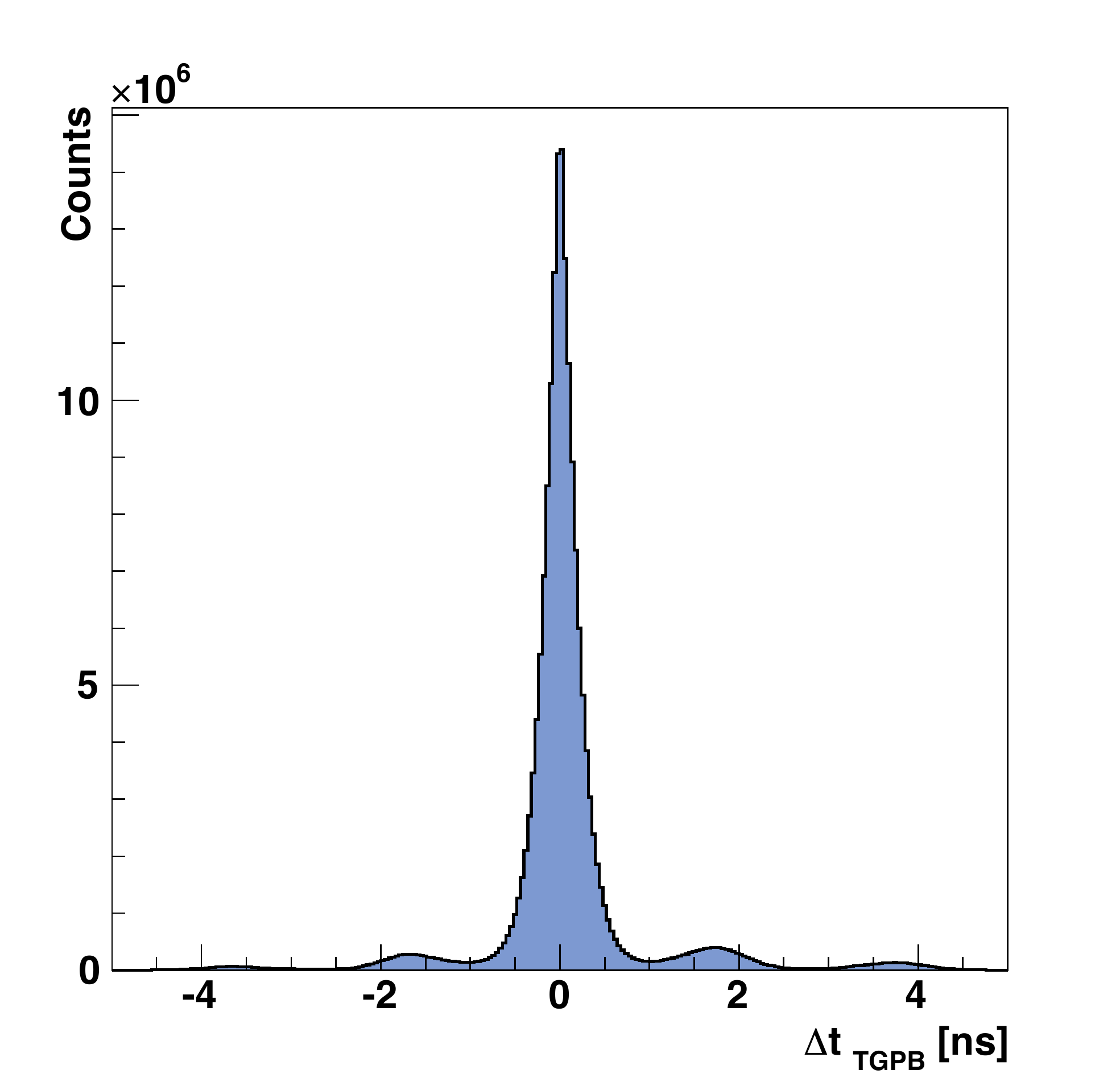}
 \caption{\label{Figure:CoincidenceTime}(Color online) Distribution of the tagger-start counter 
    coincidence times. The 2-ns bunch structure is visible. Events were considered for further 
    analysis only if a single photon candidate remained after a timing cut of 
    $|\Delta t_{\rm \,TGPB}| < 1$~ns.}
\end{figure}

The CEBAF electrons were delivered to the CLAS-g12 experiment in 2-ns bunches. Several bunches
arrived at the tagger within the trigger coincidence window and each bunch contained many electrons. 
Therefore, many photon candidates were recorded for each event; random hits could also occur from
background sources, e.g., cosmic radiation. To determine the correct initial-state photon, which 
triggered the hadronic reaction at the event vertex in the liquid-hydrogen target, the time differences 
were used between the event vertex-time based on the final-state tracks and the tagger vertex-time for 
each photon candidate. 

The event vertex-time, $t_{\rm \,event}$, was given as an average over the event's track times
\begin{eqnarray}
 t_{\rm \,track}\,=\,t_{\rm \,ST}\,-\,\frac{d}{c\,\beta_{\,m}}\,,
\end{eqnarray}
where $t_{\rm \,ST}$ denotes the start-counter time and $d$ is the distance from the interaction
point to the corresponding start-counter paddle. The time, $t_\gamma$, for each photon candidate 
is given by the recorded electron-triggered tagger time corrected for the propagation from the 
center of the liquid-hydrogen target to the event vertex along the beam axis. 
Figure~\ref{Figure:CoincidenceTime} shows the coincidence time 
$\Delta t_{\rm \,TGPB} = t_{\rm \,event} - t_\gamma$. The 2-ns time structure is clearly visible. 
In the CLAS-g12 experiment, selecting photons from the central coincidence peak and discarding events 
with more than one photon candidate resulted in a remaining non-negligible accidental background of about 
13\,\% due to the relatively high electron beam current of 60--65~nA.

\subsection{\label{Subsection:TargetPosition}The liquid-hydrogen target}
\begin{figure}[b]
 \includegraphics[width=0.48\textwidth]{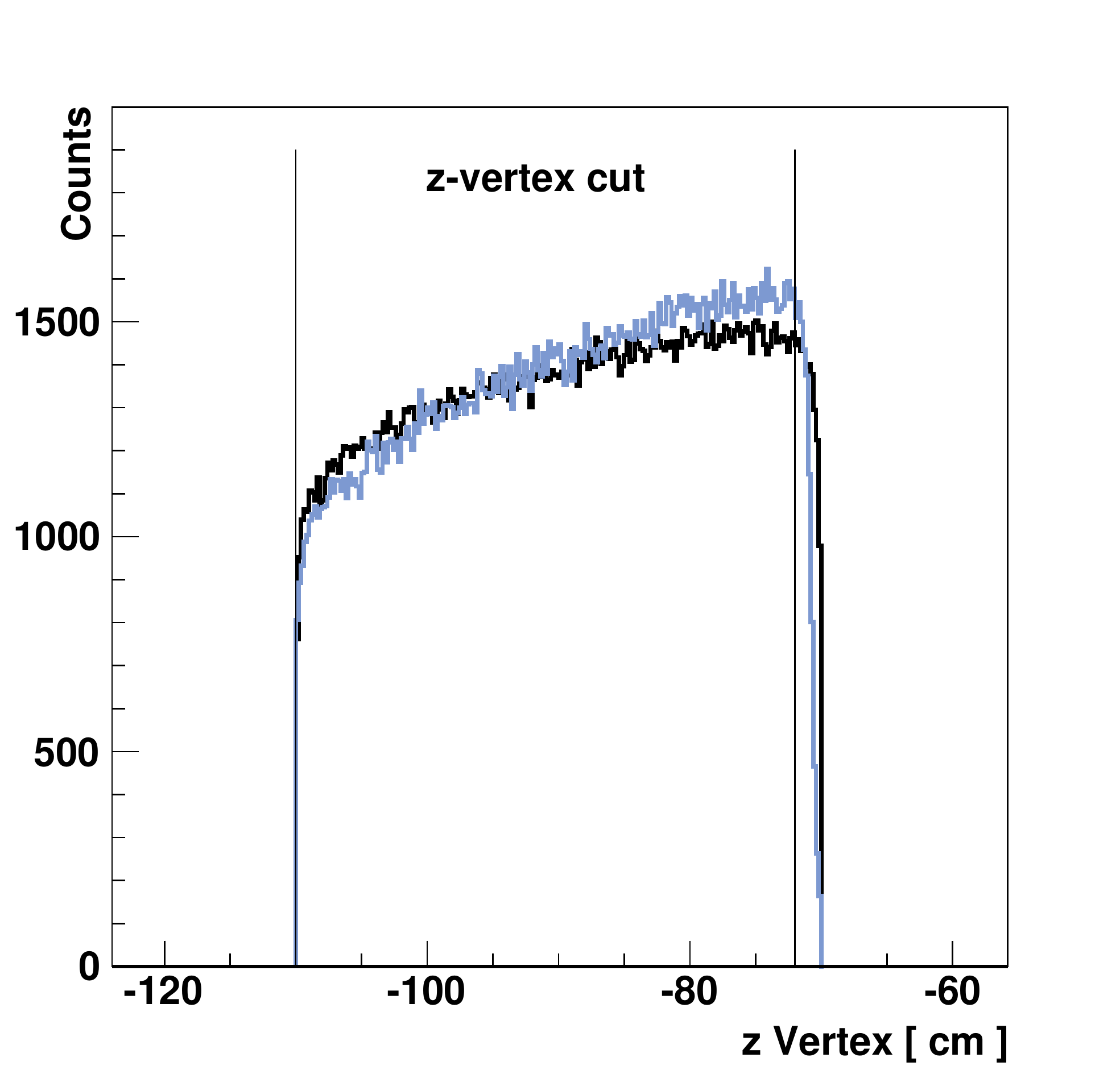}
 \caption{\label{Figure:VertexDistribution}(Color online) The $z$-vertex distribution of $\gamma p\to
   p\eta$ Monte Carlo (black) and data events (blue). The target length of 40~cm is clearly visible.
   In this experiment, the target cell was moved upstream from the CLAS center by 90~cm. The vertical
   lines define the range of the $z$-vertex cut.}
\end{figure}

In the CLAS-g12 experiment, the liquid-hydrogen target was not positioned at the center of CLAS
but was moved upstream by 90~cm to allow for the enhanced detection of peripherally produced mesons 
off the proton with the goal to search for and study excited mesons at the highest available CEBAF 
energies. Peripheral reactions are characterized by small values of the exchanged four-momentum $-t$ and 
are strongly forward peaked at high energies since the cross section is almost exponentially falling with~$t$. 
The target cell was 40~cm in length and 2~cm in diameter. The $z$-vertex distributions (coordinate along the 
beamline) for data and Monte Carlo events are shown in Fig.~\ref{Figure:VertexDistribution}. The target length 
and the position offset from the CLAS center are clearly visible.

In the CLAS-g12 experiment, the target temperature and pressure were sampled continuously 
throughout each run. Since the overall uncertainty in the target density was smaller than the 
geometrical uncertainty in the dimensions of the Kapton cell, the uncertainty in the liquid-hydrogen 
density was not considered a factor in the budget of the various systematic uncertainties.

\subsection{\label{Subsection:Trigger}Trigger}
The entire CLAS-g12 data set was classified into many different groups of runs according to their
trigger configurations. Some of these configurations applied a tagger {\it pre-scaling} to enhance 
events with high photon energies. For this analysis, we used a fraction of the total statistics 
that was not subject to pre-scaling to avoid additional complications in the absolute normalization 
of the measured angular distributions.

The TOF counters generated signals for the CLAS level-1 trigger. These detectors were positioned 
outside the CLAS tracking system in a symmetric six-sector arrangement, geometrically defined by
the coils of the CLAS toroidal magnet. For the data presented here, the trigger required a scattered
electron in the bremsstrahlung tagger in coincidence with either (a) (at least) three charged tracks 
in different sectors with no restrictions on any photon energy, or (b) only two tracks in different 
sectors with the additional requirement of observing at least one tagger photon with an energy above 
3.6~GeV. Along with several ancillary trigger conditions, these requirements resulted in a live time 
of the data-acquisition system of about 87\,\%. About 20--30 recorded photons per event were observed 
using a trigger coincidence window of approximately 100~ns.

\section{\label{Section:Selection}Calibration and event reconstruction}
The calibration of the individual spectrometer components followed the CLAS standard procedures
\cite{CLAS-NOTE-2017-002}. In the process, inefficient TOF paddles were identified and later removed 
from the analysis in a standardized approach for real data and simulated events. The latter is particularly 
important for the trigger simulation. The details of the Monte Carlo simulations are described in 
Sec.~\ref{Subsection:MonteCarlo}.

\begin{figure*}[t]
 \includegraphics[width=0.30\textwidth]{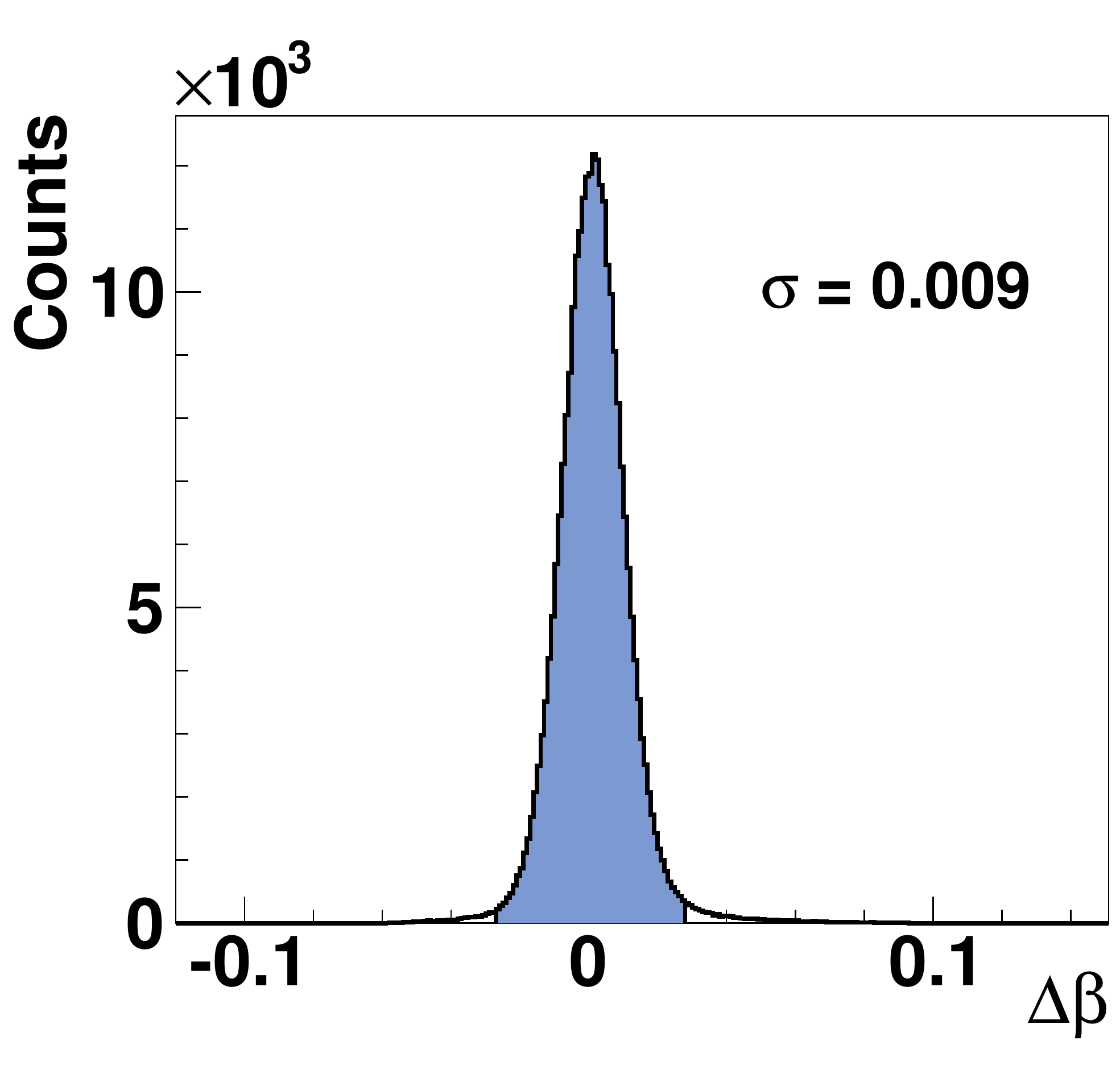}\qquad
 \includegraphics[width=0.30\textwidth]{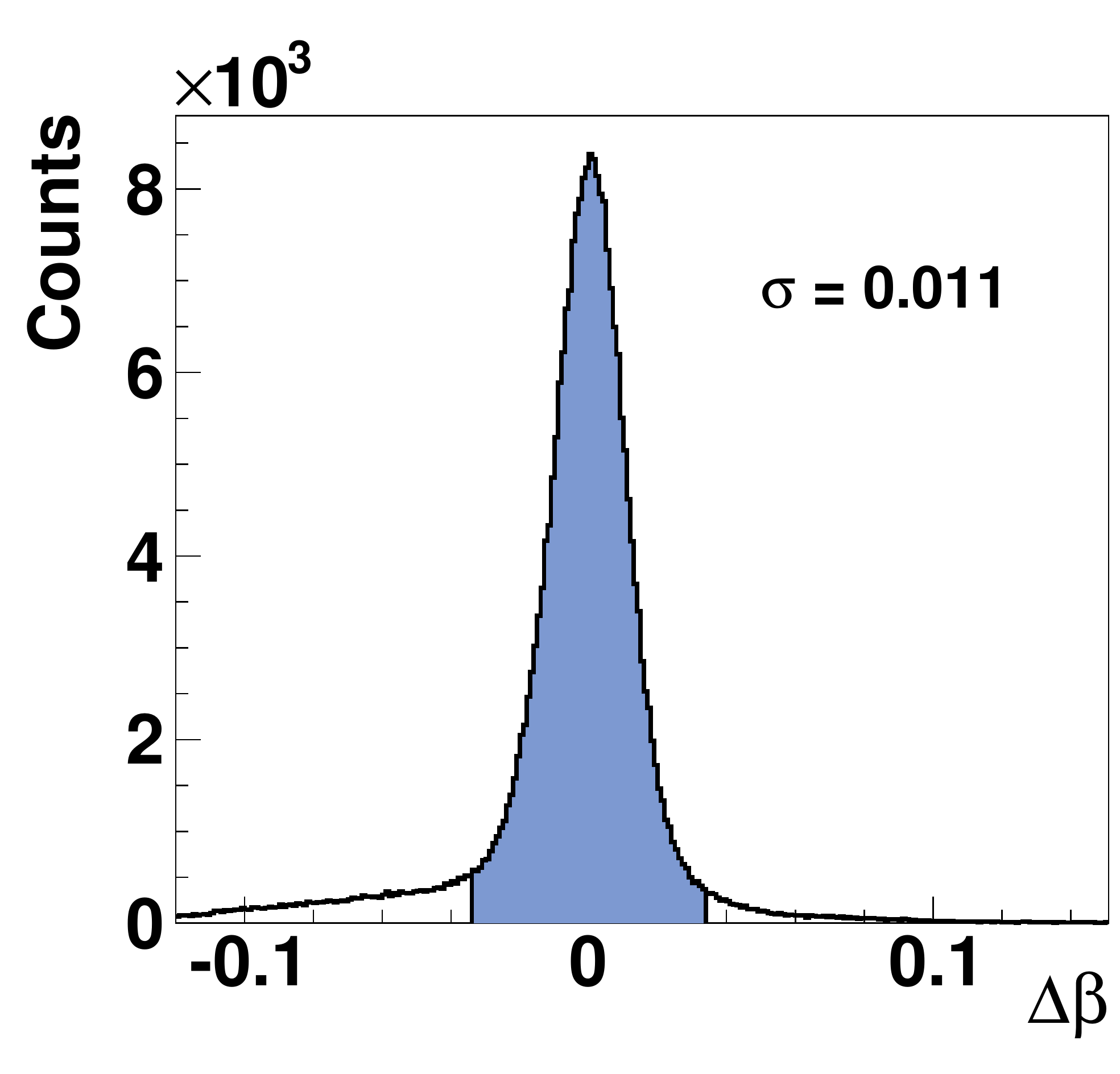}\qquad
 \includegraphics[width=0.30\textwidth]{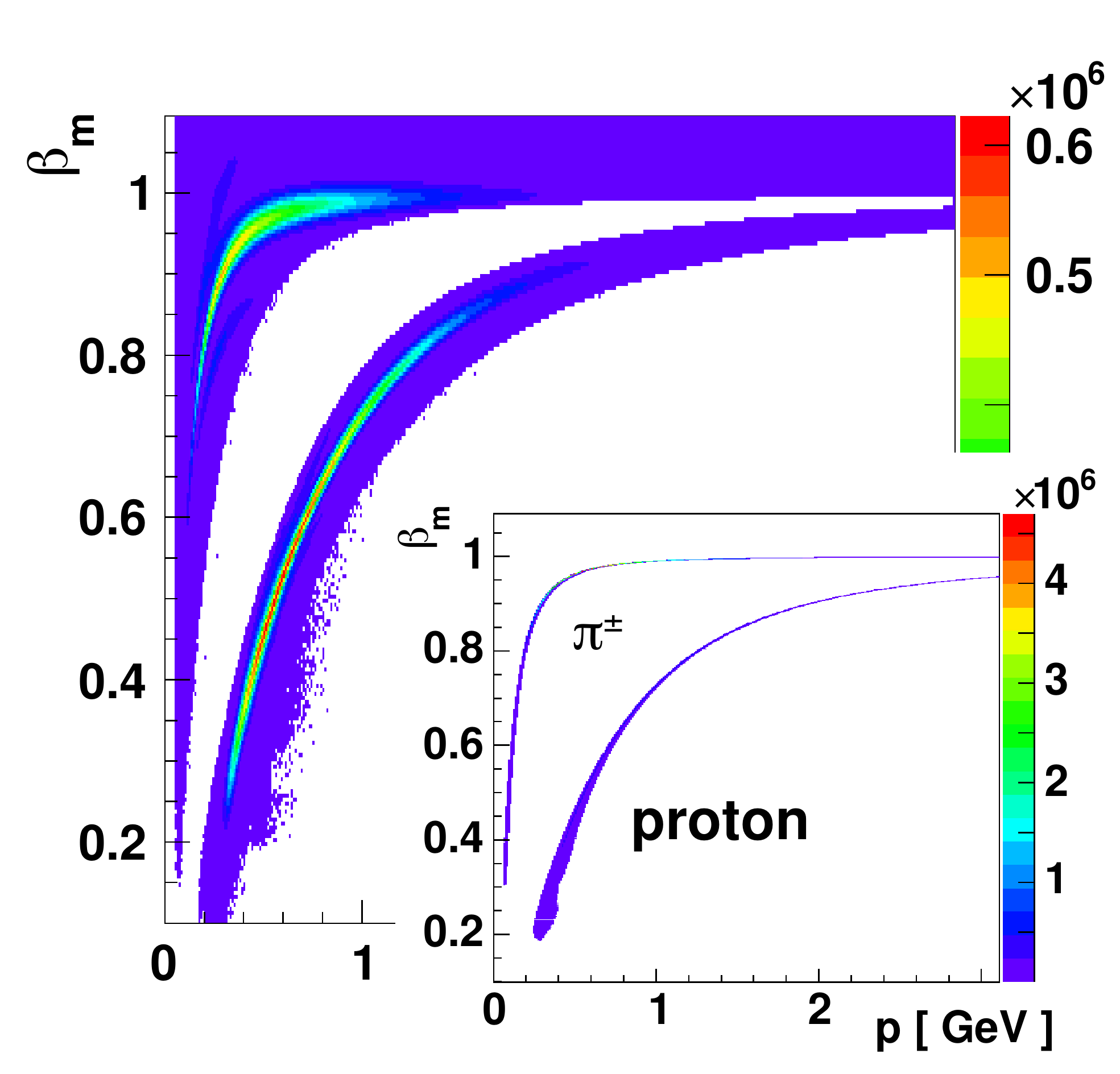}
 \caption{\label{Figure:Delta_beta} (Color online) Left and middle:
   $\Delta\beta\,=\,|\,\beta_c\,-\,\beta_m|$ distributions for protons and positively charged pions,
   respectively. The blue area indicates the $3\sigma$~cut according to Eq.~(\ref{Equation:Delta_beta}).
   Right: The distribution of $\beta_m$ vs. particle momentum before and after (inset) the $3\sigma$~cut. 
   Note that the momentum range in the inset is limited to $p < 3.1$~GeV to better illustrate the separation 
   of the two bands at low momenta.}
\end{figure*}

Charged particles emerging from the event vertex interact with various detector components and materials, 
e.g., target, beam pipe, and start counter, and therefore, are subject to energy loss along their trajectories. 
A standard CLAS software package~\cite{CLAS-NOTE-2007-016} was applied to account for these interactions. The 
CLAS drift chambers existed as three drift-chamber regions, which were located at three positions in the radial 
direction~\cite{Mestayer:2000we}. Therefore, each particle track also needed to be corrected for momentum owing 
to small misalignments of these three regions and fluctuations in the toroidal magnetic field. The momentum 
corrections for each charged particle were determined in kinematic fitting for the exclusive 
$\gamma p\to p\pi^+\pi^-$ reaction, where the mean values of the corresponding momentum pull distributions were 
tuned in an iterative procedure. The corrections were small and typically of the order of a few MeV. The set of 
simulated events did not undergo any momentum corrections.

\subsection{\label{Subsection:Preparation}Preparation of the $\pi^+\pi^-\pi^0$ final state}
The reconstruction of the $p\eta$~channel was based on preparing a data set of photoproduced 
$p\pi^+\pi^-\pi^0$ events. The same data set was also used to extract the cross sections for the reactions 
$\gamma p\to p\omega\to p\,(\pi^+\pi^-\pi^0)_{\,\omega}$ and $\gamma p\to K^0\,\Sigma^+\to 
(\pi^+\pi^-)_{\,K^0}\,(p\pi^0)_{\,\Sigma^+}$, which will be discussed in subsequent publications. The only 
major difference in extracting the cross sections for these three reactions was the subtraction of background 
events. For this reason, this section will focus on the reconstruction of the general reaction 
$\gamma p\to p\,\pi^+\pi^-\pi^0$, followed by a separate section on describing the background subtraction. The 
preparation of the final $p\eta$~event sample resulted in the reconstruction of $\approx 293\,000$~$\eta\to 
\pi^+\pi^-\pi^0$ signal events for the incident photon energy range $1.18  < E_\gamma < 4.72$~GeV or 
$1.76 < W < 3.12$~GeV in center-of-mass energy. Note that 4.72~GeV corresponds to 82.6\,\% of 
$E_{\rm \,e^-} = 5.715$~GeV and is the highest incident-photon energy that just provides sufficient statistics 
for this analysis.

\subsection*{\label{Subsection:Selection}Event reconstruction and selection criteria}
The CLAS spectrometer was optimized for detecting and measuring charged particles. However, the 
overconstrained event kinematics allows for the reconstruction of a single neutral meson. The reaction 
$\gamma p\to p\pi^+\pi^-\,(\pi^0)$ with a missing~$\pi^0$ was identified in a first step by requiring 
exactly one proton track and two charged-pion tracks. Positively and negatively charged pions were 
distinguished by their track curvatures in the toroidal field. The acceptance of $\pi^-$~mesons was 
smaller than for $\pi^+$~mesons since negatively charged tracks were bent toward the beamline and a 
large fraction escaped through the forward hole of the CLAS spectrometer. The $\pi^0$~meson was later 
identified in kinematic fitting.

Standard particle identification was then improved by evaluating $\Delta\beta$~distributions and 
applying a $3\sigma$~cut on either the proton or the $\pi^+$~meson:
\begin{equation}
\Delta\beta\,=\,|\,\beta_c\,-\,\beta_m|\,=\,\left\vert\,\sqrt{\frac{p^2}{m^2\,+\,p^2}}\,-\,\beta_m
\,\right\vert\,<\,3\sigma\,,
\label{Equation:Delta_beta}
\end{equation}
where $\beta_{\,m}$ and $\beta_{\,c}$ are based on information from the TOF and the drift-chamber system, 
respectively, as defined in Sec.~\ref{Subsection:PID}. While the quantity $\Delta\beta$ depends on 
particle momentum, the $\Delta\beta$~distribution is approximately Gaussian when summed over all 
$\beta_m$~values, with width $\sigma = 0.009$ and $0.011$ for the proton and pions, respectively. 
Figure~\ref{Figure:Delta_beta} shows the $\Delta\beta$~distributions for protons (left) and charged 
pions (center). The tail on the left side of the $\Delta\beta$~peak for pions originates from misidentified 
electrons. This small lepton contamination is not a concern since these events did not pass the kinematic 
fitter, which is described below. Also shown in Fig.~\ref{Figure:Delta_beta} (right) is the distribution of 
$\beta_m$ versus particle momentum before and after (inset) the $3\sigma$~cut according to 
Eq.~(\ref{Equation:Delta_beta}). Clear bands for the proton and the pions are visible. 

Standard fiducial cuts~\cite{CLAS-NOTE-2017-002} geometrically suppressed events outside of the active
detector regions where the acceptance was well behaved and reliably reproduced in simulations. For example, 
the magnetic field varied rapidly close to the torus coils rendering these regions difficult to simulate. This
effect was more dramatic in the forward direction, where the coils occupied a larger amount of the solid angle 
for small polar angles. Such regions were studied for charged hadrons with exclusive $\gamma p\to p\pi^+\pi^-$ 
events and defined as upper and lower limits of the azimuthal angle $\phi_{\rm \,lab}$ from the center of a 
given sector. Due to the hyperbolic geometry of CLAS and the presence of a toroidal magnetic field, the fiducial 
boundaries of $\phi_{\rm \,lab}$ are functions of a track's momentum, charge, and polar angle. Moreover, events 
were removed from this analysis if the primary interaction $z$-vertex was very close to the downstream boundary 
of the liquid-hydrogen target. The $z$-vertex resolution was dependent on the track angle and best for tracks 
that were perpendicular to the beam axis. In this experiment, the upstream shift of the long target cell from 
the center of the CLAS spectrometer affected the reconstructed upstream and downstream edges of the $z$-vertex 
distribution differently. Figure~\ref{Figure:VertexDistribution} shows that the downstream region could not be 
sufficiently well reproduced in the Monte Carlo simulations. Therefore, a cut of $-110~{\rm cm} < z$-vertex 
$< -72~{\rm cm}$ was applied to the final event sample.

The exclusive $p\pi^+\pi^-$ channel was identified as a dominant background source. This charged double-pion 
reaction has a significantly larger cross section than any other competing reaction leading to an additional 
$\pi^0$~meson in the final state. In this analysis, $p\pi^+\pi^-$ leakage into the selected 
$p\pi^+\pi^-\,(\pi^0)$ data sample was observed due to the relatively small difference in the missing masses 
of these two final states. If an incorrect initial-state photon candidate was selected with an energy higher 
than the correct incident photon, then this additional energy and $z$-momentum would allow for the reconstruction 
of an artificial $\pi^0$ in the final state that would move along the incident photon-beam direction. Therefore, 
leakage from the $\gamma p\to p\pi^+\pi^-$ channel was observed as an excess of $\pi^0$~mesons in the very
forward direction. To reduce the contribution from $\gamma p\to p\pi^+\pi^-$ background, only events with 
cos\,$\theta_{\rm c.m.}^{\,\pi^0} < 0.99$ were retained for further analysis.

\begin{figure*}[t]
  \includegraphics[width=0.32\textwidth]{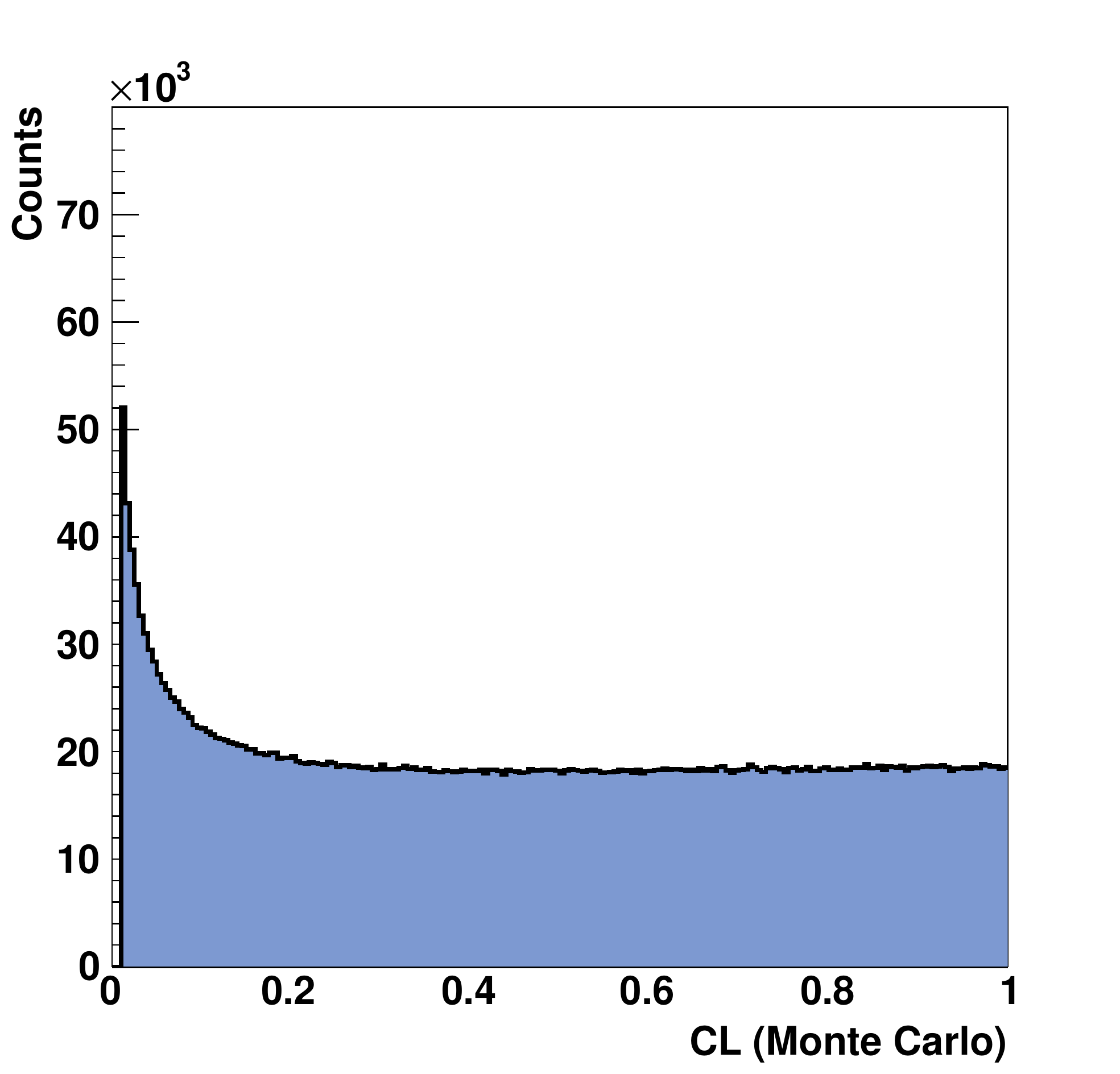}
  \includegraphics[width=0.32\textwidth]{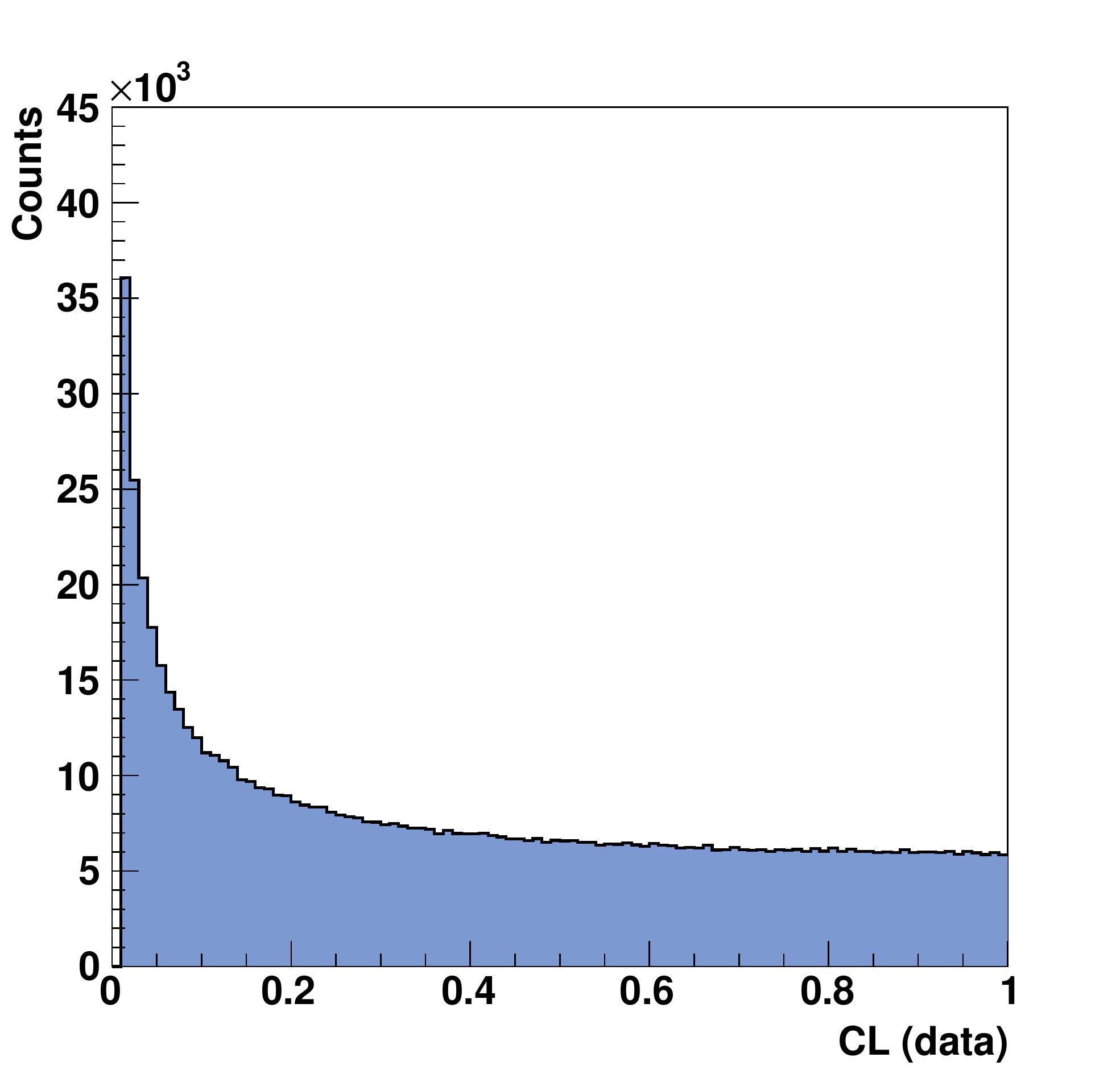}
  \includegraphics[width=0.32\textwidth]{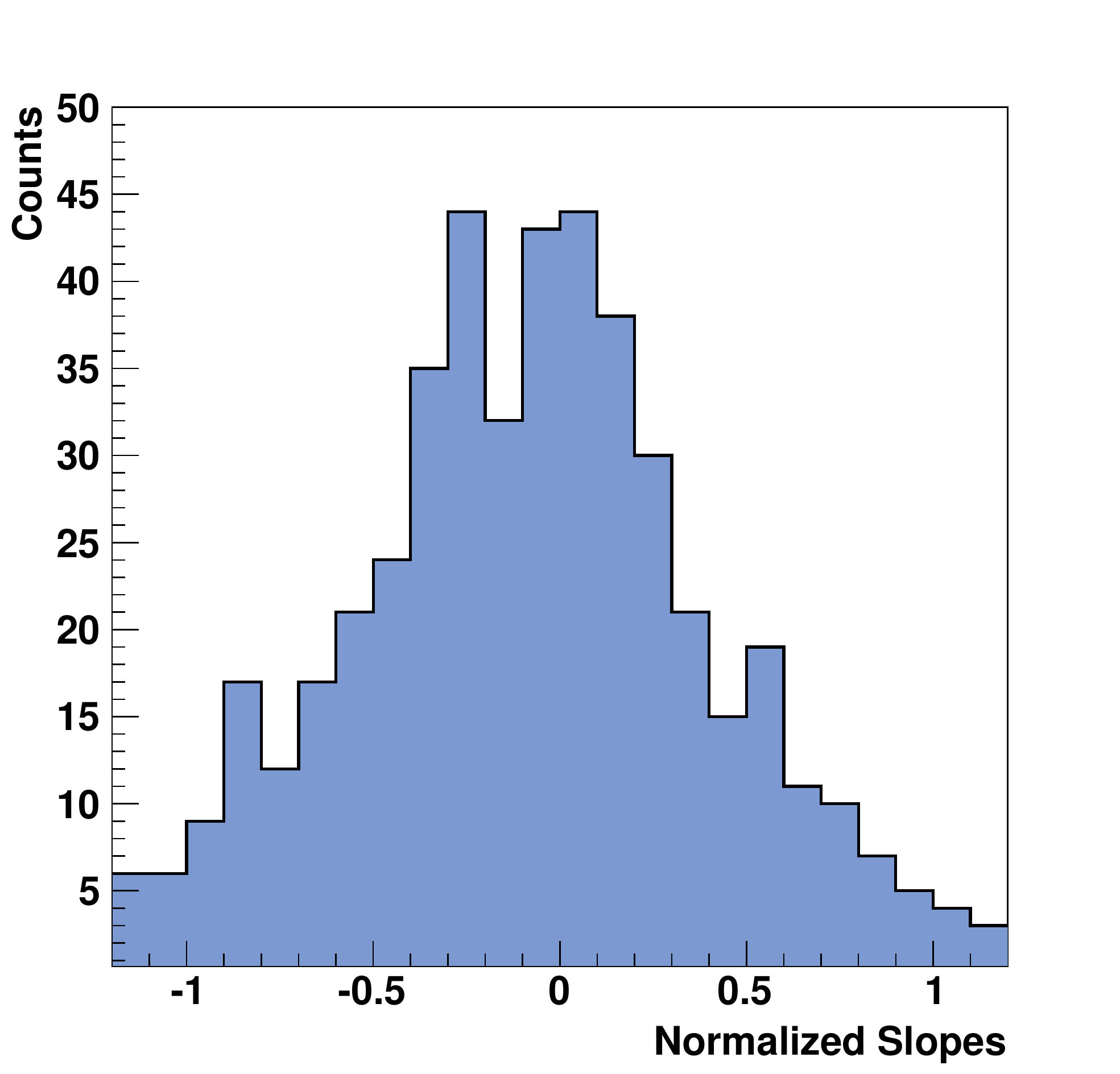}
  \caption{\label{Figure:ConfidenceLevel} (Color online) Confidence-level distribution for the
    missing-$\pi^0$ hypothesis after all corrections for Monte Carlo (MC) events (left) and CLAS-g12 data
    (center). The covariance matrix for both data and MC events was initially tuned using fully exclusive
    $\gamma p\to p\,\pi^+\pi^-$~events. Right: Distribution of normalized slopes for data events, see
    text for more details.}
\end{figure*}

In a final step, all events were subject to kinematic fitting. Events were tested separately for energy 
and momentum conservation in a four-constraint (4C) fit to identify the reaction $\gamma p\to p\pi^+\pi^-$ 
and in a one-constraint (1C) fit to test for a missing~$\pi^0$. Three pieces of information are needed for 
a missing particle. The missing four-momentum initially introduces four additional unknown parameters. However, 
the particle's energy and momentum are related by the invariant mass. Since energy and momentum conservation 
provides four constraints, the missing-particle hypothesis reduces to a one-constraint fit. The exclusive 
reaction $\gamma p\to p\pi^+\pi^-$ was used to tune the covariance matrix with a set of common parameters 
that have also been applied in other CLAS-g12 analyses involving kinematic fitting. This procedure secures 
Gaussian pull distributions and a flat confidence-level (CL) distribution, where the CL~denotes the 
{\it goodness of fit} of the statistical model applied to the data and is defined as the integral over the 
$\chi^2$~probability density function in the range $[\,\chi^2,\infty\,]$~\cite{brandt}. 

\begin{figure}[b]
 \includegraphics[width=0.46\textwidth]{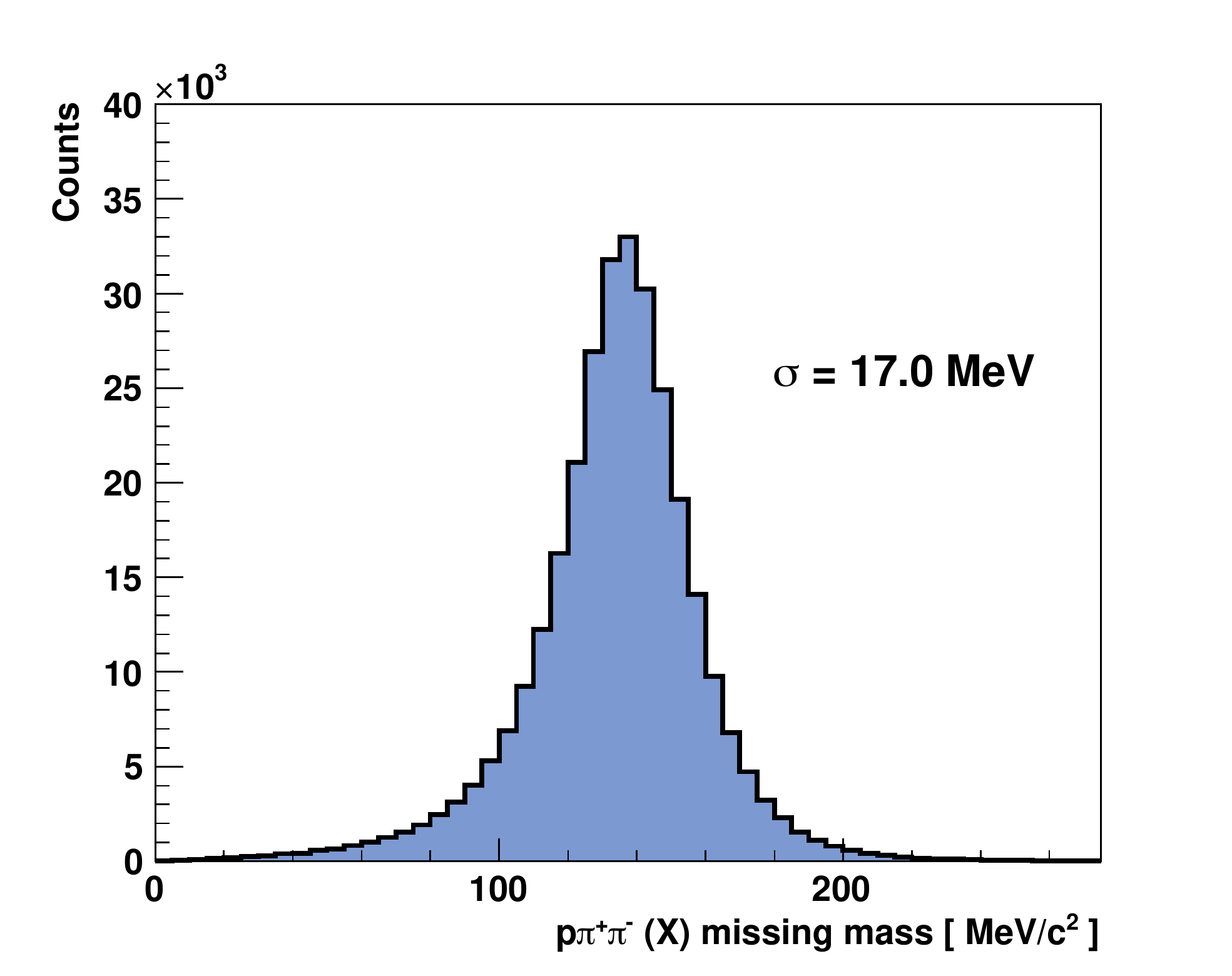}
 \caption{\label{Figure:MissingMass}(Color online) Missing mass~$X$ in the reaction $\gamma p\to
    p\eta\to p\,\pi^+\pi^-\,(X)$ after all cuts and the final $\eta$~background subtraction. A clean
    $\pi^0$~peak is visible.}
\end{figure}

Figure~\ref{Figure:ConfidenceLevel} (center) shows the confidence-level distribution for the missing-$\pi^0$ 
hypothesis after all corrections; the distribution is fairly flat. In addition to the quality of the global 
CL and pull~distributions, the flat shape of the CL~distributions was also checked in all relevant kinematic 
regions by considering the {\it normalized slope} of each distribution:
\begin{eqnarray}
\bar{a} = \frac{a}{a/2 + b}\,,
\end{eqnarray}
where $a$ is the slope and $b$ is the $y$~intercept obtained by fitting a first-order polynomial to the 
confidence-level distribution on the intervall [\,0.6,\,1.0\,]. Figure~\ref{Figure:ConfidenceLevel} (right) 
shows the respective normalized slopes integrated over all analyzed energies and $\eta$~center-of-mass angles. 
The distribution is symmetric and centered at {\it zero} demonstrating the relative flatness of the 
CL~distributions in all kinematic bins and thus, the good understanding of the measurement uncertainties. 
Events in this analysis were retained with a confidence-level cut of $p > 0.01$.

Figure~\ref{Figure:MissingMass} shows the missing mass using non-kinematically fitted four-vectors in the 
reaction $\gamma p\to p\eta\to p\,\pi^+\pi^-\,(X)$, integrated over all available incident photon energies, 
after all cuts and the final $\eta$~background subtraction described in the following section. A clean 
$\pi^0$~peak is visible with a Gaussian width of $\sigma = 17.0$~MeV.

\subsection*{\label{Subsection:MonteCarlo}Monte Carlo simulations}
The performance of the experimental setup was studied in GEANT3-based~\cite{Brun:1994aa} Monte-Carlo 
(MC) simulations. The acceptance for the reaction $\gamma p\to p\eta\to p\,\pi^+\pi^-\pi^0$ was 
determined by generating events, which were evenly distributed across the available phase space. 
The MC events were then analyzed using the same reconstruction and selection criteria, which were 
applied to the measured data events. The simulated tracks were corrected for the energy loss along 
their trajectories but were not subject to any momentum corrections since all the DC components
were perfectly positioned in the simulations and a homogeneous magnetic field was used. The same 
hypotheses were tested in the kinematic fits and events selected with the same confidence level 
cut. The acceptance for each kinematic bin was then defined as the ratio of the number of generated
to reconstructed MC events:
\begin{eqnarray}
A_{\,\gamma p\,\to\, p\eta} \,=\,\frac{N_{\rm \,rec,\,MC}}{N_{\rm \,gen,\,MC}}~.
\end{eqnarray}

In the real CLAS-g12 data, information about the trigger condition was encoded in the so-called trigger word, 
which was available in the data stream for every event. The overall trigger (in)efficiency was evaluated by 
studying the efficiency of individual charged tracks to produce a trigger-level signal in a sample of exclusive 
$\gamma p\to p\,\pi^+\pi^-$ events, where each final-state particle was detected in a different sector of CLAS. 
Since one of the trigger conditions required only (at least) two charged tracks in two different sectors (see 
Sec.~\ref{Subsection:Trigger}), any inefficiencies could be studied by comparing with the encoded trigger 
information. The trigger efficiency for a charged track was then given as the fraction of events where a third 
particle could be reconstructed in a different sector but the information was not recorded in the corresponding 
trigger bit. Trigger efficiency maps were developed for each particle type (proton, $\pi^+$, $\pi^-$) as a 
function of sector ID, TOF counter, and azimuthal angle, $\phi$. The average CLAS-g12 efficiency values for the 
proton, $\pi^+$, and $\pi^-$ are 0.89, 0.83, and 0.75, respectively. These maps were applied in the Monte Carlo 
simulations by generating a random number for each track and removing the track if the random number exceeded 
the corresponding efficiency stored in the map.

\subsection{\label{Subsection:Background}Background subtraction}
In the determination of the $\eta$~photoproduction cross sections reported here, non-signal background
events were removed in a probabilistic event-based approach called the ``$Q$-factor method,'' which is fully
described in Ref.~\cite{Williams:2008sh}. A brief summary of the method and its application to the data from
CLAS-g12 is given in this section.

\begin{figure}[b]
 \includegraphics[width=0.5\textwidth]{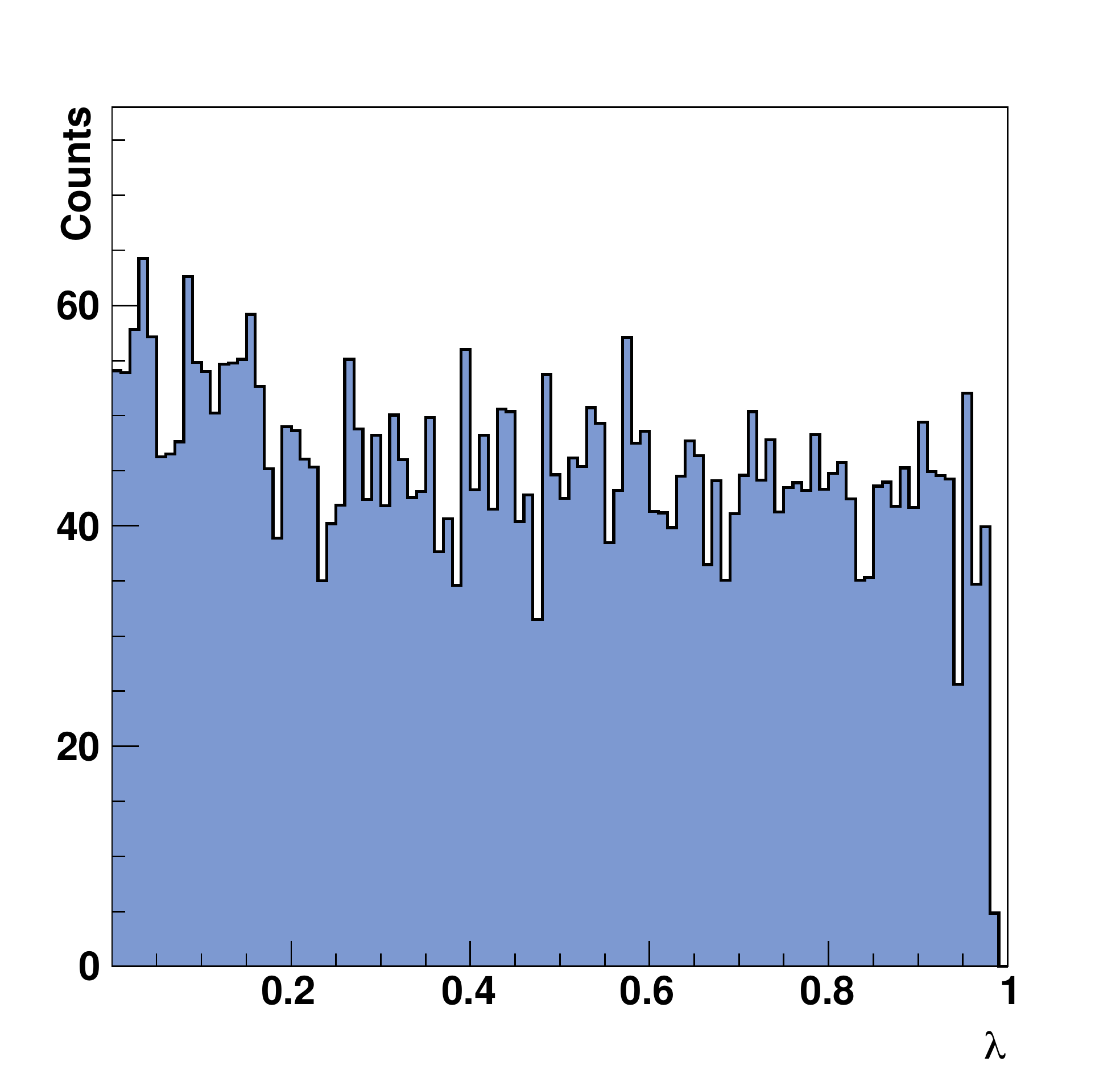}
 \caption{\label{Figure:lambda} (Color online) Typical example of a normalized
   $\lambda = |\,\vec{p}_{\pi^+}\,\times\,\vec{p}_{\pi^-}|^2$ distribution for the center-of-mass
   energy bin $W\in [\,2360,\,2400\,]$~MeV.}
\end{figure}

\begin{figure*}[t]
 \includegraphics[width=1.0\textwidth,height=0.25\textheight]{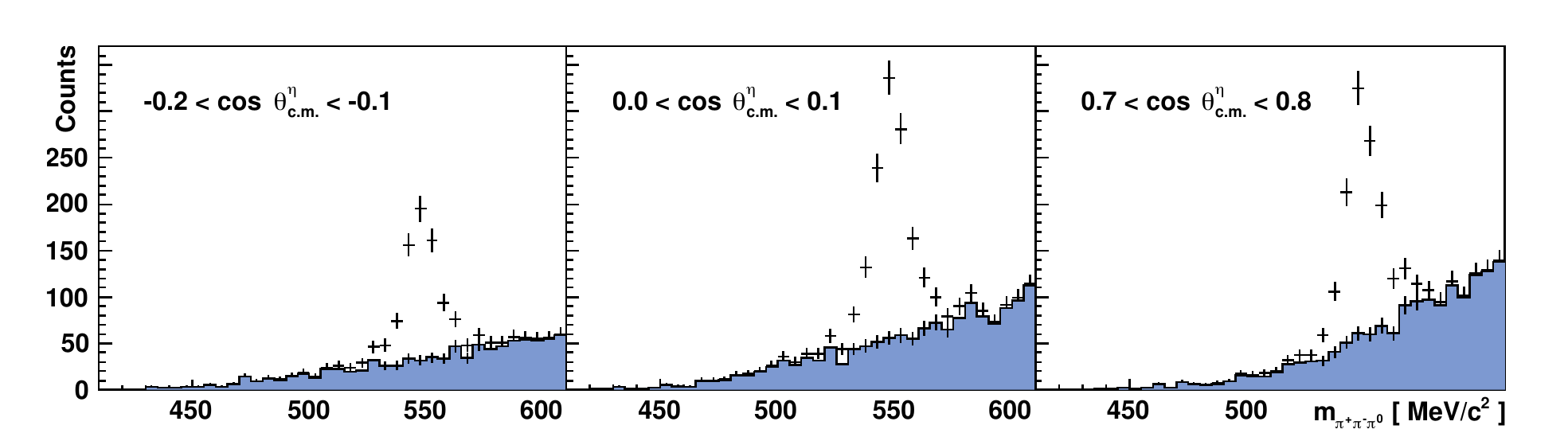}
 \caption{\label{Figure:MassDistributionsEta}(Color online) Examples of $\pi^+\pi^-\pi^0$~mass
   distributions for the center-of-mass energy range $W\in [\,1.90,\,1.92\,]$~GeV for events that were
   subject to the $Q$-factor fitting (background subtraction). These events survived all kinematic
   cuts. The invariant $3\pi$~mass of each event weighted by $1-Q$ gives the blue area (background),
   whereas the signal peak comes from the invariant mass weighted by~$Q$.}
\end{figure*}

For every event in this analysis, a quality factor (or $Q$~value) was determined that describes the probability
for an event to be a signal event as opposed to background. The approach used the unbinned maximum-likelihood
technique. For every selected $\gamma p\to p\eta$~event and its $N_c$ kinematically nearest neighbors, the
following function was fit to the invariant $M_{\pi^+\pi^-\pi^0}$~mass distribution:
\begin{equation}
  f(x) \,=\, N\,\cdot [ f_{s}\,\cdot\, S(x) \,+\, ( 1 \,-\, f_{s} )\,\cdot\, B(x) ]\,,
  \label{Equation:Q-Function}
\end{equation}
where $S(x)$ and $B(x)$ denote the signal and the background probability density functions, respectively, and
$x = M_{\pi^+\pi^-\pi^0}$. A double-Gaussian profile was chosen for the signal and the background shape was 
modeled with a second-order Chebyshev polynomial. The parameter~$N$ in Eq.~(\ref{Equation:Q-Function}) is a 
normalization constant and $f_{s}$ is the signal fraction with a value between 0 and 1.

The kinematically nearest neighbor events were selected by defining a distance metric for the phase space
spanned by a set of kinematic variables $O_k$. These independent quantities were chosen to be
\begin{equation}
{\rm cos}\,\theta^{\,\eta}_{\rm c.m.},~{\rm cos}\,\theta_{\rm \,HEL},~\phi_{\rm \,HEL},~\phi^{\,\eta}_{\rm lab},~\lambda\,,
\label{Equation:Quantities}
\end{equation}
where cos$\,\theta^{\,\eta}_{\rm c.m.}$ denotes the cosine of the polar angle of the $\eta$ in the center-of-mass
frame, cos$\,\theta_{\rm \,HEL}$ and $\phi_{\rm \,HEL}$ describe the orientation of the $\eta$~decay plane in the 
helicity frame, and $\phi^{\,\eta}_{\rm lab }$ is the azimuthal angle of the $\eta$ in the laboratory frame. The 
variable $\lambda = |\,\vec{p}_{\pi^+}\,\times\,\vec{p}_{\pi^-}|^2\,/\,\lambda_{\rm \,max}$ is defined in terms 
of the pion momenta in the $\eta$~rest frame and is proportional to the $\eta\to\pi^+\pi^-\pi^0$ decay amplitude 
as a consequence of isospin conservation~\cite{Williams:2009ab}, with $\lambda_{\rm \,max}$ defined 
as~\cite{Weidenauer:1993mv}
\begin{equation}
\lambda_{\rm \,max} \,=\, K^2\,\bigg(\frac{K^2}{108}\,+\,\frac{mK}{9}\,+\,\frac{m^2}{3} \bigg)\,,
\end{equation}
for a totally symmetric decay, where $K = T_1 + T_2 + T_3$ is the sum of the $\pi^{\pm,\,0}$~kinetic energies 
and $m$ is the $\pi^\pm$~mass. 

Initially defined for vector mesons, $\lambda$ has a limited physics interpretation for pseudoscalar mesons but 
still serves as an independent kinematic variable in this analysis. The background subtraction described in this 
section was performed simultaneously for the $\omega$~and $\eta$~meson decaying to the same  $\pi^+\pi^-\pi^0$ 
final state. Results on cross section measurements for $\gamma p\to p\omega$ will be presented in a forthcoming 
publication~\cite{Akbar:2020}. The parameter $\lambda$ varies between 0 and 1 and the number of events as a 
function of $\lambda$ shows a linearly increasing behavior for vector mesons, whereas a flat distribution is 
expected for the $\eta$~meson. This is nicely observed in Fig.~\ref{Figure:lambda}. Using the quantities listed
in Eq.~(\ref{Equation:Quantities}), the kinematic distance between two events $i$ and $j$ is defined as
\begin{equation}
  d_{ij}^2 \,=\,\sum_{k=1}^5 \bigg( \frac{O^i_k\,-\,O^j_k}{\Delta_k}\bigg)^2\,,
\label{Equation:Distance}
\end{equation}
where the $O_k$ denotes the set of kinematic variables for the two events $i$ and $j$, and $\Delta_k$ is the 
full range for the kinematic variable $k$.

The $Q$~value for a selected $\gamma p\to p\eta$~event is finally given as the signal component at the event's
invariant $\pi^+\pi^-\pi^0$~mass in the overall mass distribution of the event and its $N_c = 500$~nearest 
neighbors:
\begin{equation}
  Q \,=\, \frac{s(x)}{s(x) \,+\, b(x)}\,,
  \label{equ:Q_factor}
\end{equation}
where $x$ is again the invariant mass of the $\pi^+\pi^-\pi^0$~system, $s(x) = f_{s} \cdot S(x)$, and 
$b(x) = (1-f_{s}) \cdot B(x)$ [see also Eq.~(\ref{Equation:Q-Function})].

\begin{figure}[b]
 \includegraphics[width=0.48\textwidth]{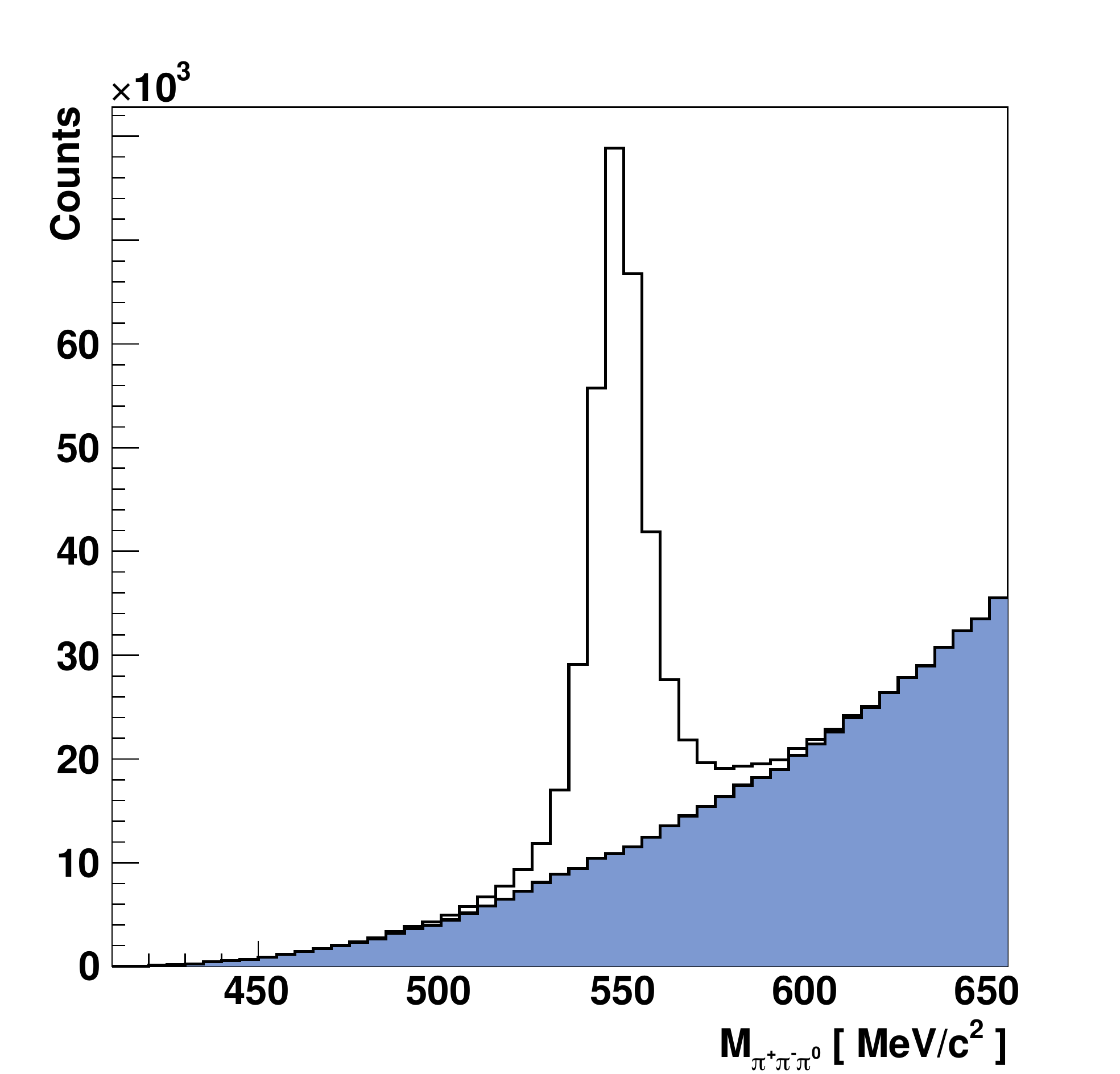}
 \caption{\label{Figure:TotalMass} (Color online) Total invariant $\pi^+\pi^-\pi^0$~mass distribution
    for the center-of-mass energy $W\in [\,1.76,\,2.36\,]$~GeV corresponding to the combined
    $\gamma p\to p\eta$~event statistics of Figs.~\ref{Figure:W-1760-1880}--\ref{Figure:W-2120-2360}.}
\end{figure}

The $Q$~values were then used as weight factors for various kinematic distributions in this analysis.
Figure~\ref{Figure:MassDistributionsEta} shows examples of the resulting separation of signal and background
in the invariant $\pi^+\pi^-\pi^0$ mass distribution. Three angle bins are presented in the energy range 
$W\in [\,1.90,\,1.92\,]$~GeV. The sum of the signal (white area) and the background (blue area) is identical 
to the total unweighted mass distribution, whereas the invariant $3\pi$~mass of each event weighted by $1-Q$ 
gives the background alone. Figure~\ref{Figure:TotalMass} shows the total invariant $\pi^+\pi^-\pi^0$ mass 
distribution for the energy range $W\in [\,1.76,\,2.36\,]$~GeV representing the underlying event statistics 
in Figs.~\ref{Figure:W-1760-1880}--\ref{Figure:W-2120-2360}. An excellent signal\,/\,background separation 
is observed with $\approx 269\,000$~$\eta$~events in the signal peak.

\section{\label{Section:Extraction}Extraction of cross sections}
The differential cross sections, d$\sigma$/d$\Omega$, for the reaction $\gamma p\to p\eta$ are 
determined according to
\begin{equation}
\frac{\rm d\sigma}{\rm d\Omega}\,=\,\frac{N_{\rm \,\gamma p\,\to\, p\eta}}{A_{\rm \,\gamma 
p\,\to\, p\eta}}~\frac{1}{N_{\gamma}\,\rho_{\rm \,target}}~\frac{1}{\Delta\Omega}~\frac{1}{\rm BR}~,
\label{Equation:CrossSection}
\end{equation}
where\par\smallskip

%\begin{table}[H]
%\addtolength{\extrarowheight}{3pt}
\begin{tabular}{crl}
& $\rho_{\rm \,target}$\,: & target area density\\[0.6ex]
& $N_{\rm \,\gamma p\,\to\, p\eta}$\,: & number of reconstructed signal events\\
& & in a ($W$,~cos\,$\theta_{\rm \,c.m.}$) bin\\[0.6ex]
& $N_\gamma$\,: & number of photons in an incident $E_\gamma$~bin\\[0.6ex]
& $A_{\rm \,\gamma p\,\to\, p\eta}$\,: & acceptance in a 
  ($W$,~cos\,$\theta_{\rm \,c.m.}$) bin\\[0.6ex]
& $\Delta\Omega$\,: & solid-angle interval $\Delta\Omega = 2\pi\,\Delta 
  {\rm cos}\,(\theta_{\rm \,c.m.})$\\[0.6ex]
& BR\,: & decay branching fraction.
\end{tabular}
%\end{table}

The target area density, i.e., the number of atoms in the target material per cross-sectional 
area (orthogonal to the photon beam), is given by
\begin{equation}
\rho_{\rm \,target}\,=\, 2\,\frac{\rho\,({\rm H}_2)\,N_A\,L}{M_{\rm \,mol}\,({\rm H}_2)}\,=\, 
16.992\cdot 10^{-7} \mu {\rm b}^{-1}\,,
\end{equation}
where $\rho\,({\rm H}_2) = 0.0711$~g/cm$^3$~\cite{CLAS-NOTE-2017-002} is the average density, 
$M_{\rm\,mol}=2.01588$~g/mol is the molar mass of liquid~H$_2$, and $L = 40.0$~cm is the length 
of the CLAS-g12 target cell. Finally, $N_A = 6.022\cdot 10^{23}$ mol$^{-1}$ is Avogadro's number. 
The factor of two accounts for the molecular composition of hydrogen~(H$_2$).

The solid angle in steradians equals the area of a segment of a unit sphere. The full solid angle 
of a sphere measured from any point in its interior is thus $2\cdot 2\pi = 4\pi$~sr, where $2\pi$ 
originates from integrating over the azimuthal angle and the factor of two from integrating over
${\rm sin}\,\theta\,{\rm d}\theta$ (polar angle). Since the differential cross sections are integrated 
over $\phi_{\rm \,lab}$ but are binned in cos\,$\theta_{\rm \,c.m.}$, $\Delta\Omega = 2\pi\,\Delta 
{\rm cos}\,(\theta_{\rm \,c.m.})$ was used in Eq.~(\ref{Equation:CrossSection}) and $\Delta 
{\rm cos}\,(\theta_{\rm \,c.m.})$ = 2 \,/\,($\#$ of angle bins). In this analysis, the available
statistics allowed for 20 angle bins and thus, $\Delta\Omega = 0.6283$.

The branching fraction for the charged decay mode $\eta\to\pi^+\pi^-\pi^0$ of
$\Gamma_{\pi^+\pi^-\pi^0}\,/\,\Gamma = (22.92\pm 0.28)$\,\% was taken from Ref.~\cite{Tanabashi:2018oca},
where $\Gamma = (1.31\pm 0.05)$~keV~\cite{Tanabashi:2018oca}.

\subsection{\label{Subsection:Normalization}Normalization}
The photon flux for the absolute normalization of the extracted differential cross sections was determined
using standard CLAS procedures. The method is described in Ref.~\cite{CLAS-NOTE-2005-002} and based on 
comparing the number of ``good'' electrons in the tagger with the number of photons traversing the 
liquid-hydrogen target measured with a total absorption counter (TAC) placed directly in the photon beam. 
Such normalization runs were carried out at about 10\,\% of the production beam current using a thinner 
bremsstrahlung radiator to determine the tagging ratio $\epsilon^{\rm T}$ of each T-counter. The tagging 
ratio is given by the ratio of ``good'' tagger hits in coincidence with the TAC to the total number of 
``good'' hits in the tagger and is approximately between 75\,\% and 80\,\%. Photons can be lost on the way 
from the tagger to the target due to dispersion of the beam, collimation, and M\o ller scattering, for instance. 
The number of ``good'' electrons is given by integrating the observed electron rates at the tagger over the data 
acquisition (DAQ) live time of the experiment, which is measured with a clock. The number of tagged photons per 
T-counter is then given by:
\begin{eqnarray}
  N^{\rm T}_\gamma \,=\, \frac{N^{\rm T}_{\rm e^-} \,\times\, \epsilon^{\rm  T}}{1\,-\,\alpha}~,
\end{eqnarray}
where $N^{\rm T}_{\rm e^-}$ is the number ``good'' electrons per T-counter and the photon attenuation 
factor, $\alpha$, denotes the small fractional loss of photons from the liquid-hydrogen target to the TAC.

\subsection{\label{Subsection:Uncertainties}Systematic uncertainties}
The statistical uncertainties were determined from the number of $p\eta$ events in each ($W$,
cos\,$\theta^{\,\eta}_{\rm c.m.}$) or ($W$, $-t$)~bin, and are included in the uncertainties shown for 
all data points. In this analysis, the effective number of events in each kinematic bin was given by 
summing over all $Q$~values of the contributing events. 

The overall systematic uncertainty includes uncertainties in the normalization, as well as contributions 
from reconstruction-related sources and the background-subtraction method. An overview of the different 
fractional contributions (\%~uncertainties) is given in Table~\ref{Table:Uncertainties}. These contributions
are not included in the following results figures. A brief discussion of the contribution from the 
background-subtraction method is given in this section below. Such contributions are included in the 
uncertainty shown for each data point (added in quadrature to the statistical uncertainty). 

\begin{table}[b]
%\addtolength{\extrarowheight}{3pt}
\begin{tabular}{cc}
Source of Uncertainty & \%~Uncertainty\\[0.5ex]\hline
  Sector-by-sector relative acceptance~\cite{CLAS-NOTE-2017-002} & 5.9\\
  Fiducial cuts & 2.5\\
  $z$-vertex cut & 2.6\\
  Upstream\,/\,downstream target half & 1.5\\
  Kinematic fitting (CL cut) & 1.6\\
  Trigger efficiency correction & 1.1\\ 
  %Incident photon flux & 1.7\\
  Liquid-hydrogen target~\cite{CLAS-NOTE-2017-002} & 0.5\\
  Normalization (photon flux)~\cite{CLAS-NOTE-2017-002} & 5.7\\ 
  Branching fraction ($\eta\to\pi^+\pi^-\pi^0$) & 0.28
\end{tabular}
\caption{\label{Table:Uncertainties}Summary of the fractional contributions to the overall systematic
uncertainty.}
\end{table}

An individual event's $Q$~value is based on a fit to the invariant $\pi^+\pi^-\pi^0$~mass distribution that
is formed by the event and its kinematically nearest neighbors using the maximum-likelihood technique. The 
covariance matrix, $C_{\eta}$, for the set of fit parameters, $\vec{\eta}$, was used to determine the 
uncertainty of the $Q$~value for the given event:
\begin{equation}
\sigma^2_Q\,=\,\sum_{k,\,m}\,\frac{\partial Q}{\partial\eta_{k}}\,\big(C_{\eta}^{-1}\big)\,\frac{\partial Q}
{\partial \eta_{m}}~,
\end{equation}
where the summation runs over all signal and background parameters used in the total fit function $f(x)$, 
which is defined in Eq.~(\ref{Equation:Q-Function}).

\begin{figure*}[t]
 \includegraphics[width=1.0\textwidth]{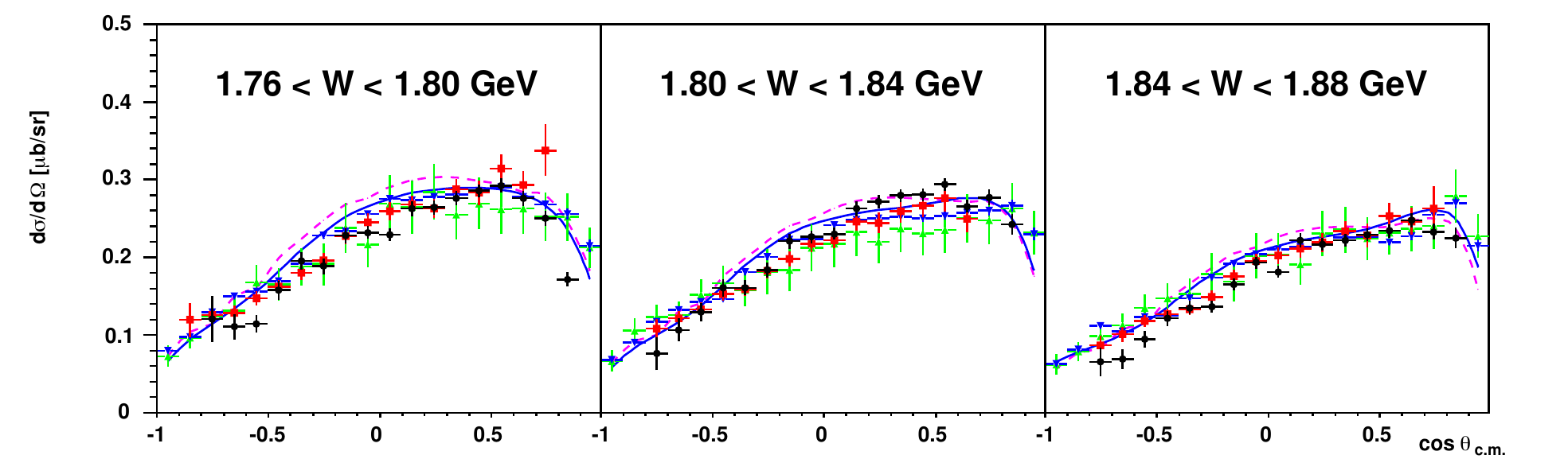}
 \caption{\label{Figure:W-1760-1880} (Color online) The differential cross sections d$\sigma$/d$\Omega$
    for three 40-MeV-wide center-of-mass energy $W$~bins. The new CLAS data are shown as the black solid
    circles ({\large $\bullet$}) and the uncertainties associated with each point comprise the statistical
    uncertainty and contributions from the $Q$-value correlation uncertainty added in quadrature.
    Also shown for comparison are data from CLAS-g11a~\cite{Williams:2009yj}~({\tiny\color{red}
    {$\blacksquare$}}), the A2 Collaboration at MAMI~\cite{Kashevarov:2017kqb}~({\color{blue}
    {$\blacktriangledown$}}) using their published center-of-bin energies of $W = 1.78$~GeV (left),
    1.82~GeV (center), 1.86~GeV (right), and the CBELSA/TAPS Collaboration at ELSA~\cite{Crede:2009zzb}
    ({\color{green}{$\blacktriangle$}}). The blue solid and purple dashed curves denote the $\eta$-MAID\,2018
    \cite{Tiator:2018heh} and the BnGa\,2019~\cite{Muller:2019qxg} description of the
    $\gamma p\to p\eta$~cross section, respectively.}
\end{figure*}

The $Q$~factor method naturally led to some correlations among events and their nearest neighbors because 
events could serve as neighbors for many {\it seed} events. The systematic ``correlation'' uncertainty of 
the $\eta$~yield in each kinematic bin due to the method as such was given by: 
\begin{equation}
\sigma^2_\eta\,=\,\sum_{i,\,j}\,\sigma_Q^i\,\rho_{ij}\,\sigma_Q^j\,,
\label{Equation:Q_error}
\end{equation}
where the sum $i,j$ was taken over all the events in the kinematic bin, $\sigma_Q^i$~and $\sigma_Q^j$ denote
the fit uncertainties for events $i$ and $j$, and $\rho_{ij}$ represents the correlation factor between events 
$i$ and $j$. The correlation factor is simply the fraction of shared nearest-neighbor events and a number 
between {\it zero} and {\it one}. In high-statistics event samples, the correlation among events is typically 
small and the corresponding contribution to the overall systematic uncertainty is negligible, whereas in 
low-statistics samples, the contribution can quickly exceed the basic statistical uncertainty.

The contribution from the $Q$-factor method was then added to the statistical uncertainty in quadrature to
obtain the total ``statistics-based'' uncertainty that is shown for each data point in subsequent figures:
\begin{equation}
\sigma^2\,=\,\sigma^2_\eta\,+\,\sigma^2_{\rm statistical}\,.
\end{equation}

An additional CL~cut of $p > 0.05$ was examined and the resulting differential cross sections were compared 
with the original results when a nominal cut of just $p > 0.01$ was used. Both the difference and ratio 
distributions were observed to be symmetric and Gaussian reflecting a change in the results, which is mostly 
statistical in nature due to the loss of events when using a larger $p$~value. A possible contribution from
the different slopes of the $z$-vertex distributions in data and Monte Carlo (observed in 
Fig.~\ref{Figure:VertexDistribution}) was studied by comparing the differential cross section results based 
on events which originated from either the upstream or downstream half of the liquid hydrogen target. The 
ratio distribution was found to be symmetric, but shifted away from unity by about 1.5\,\%. Finally, this
analysis was also repeated without any $z$-vertex cut. A Gaussian fit to the corresponding (very narrow and 
symmetric) ratio distribution resulted in a small shift of 2.6\,\%.

In the ideal CLAS-g12 experiment, the detector response will be the same in each of the six CLAS sectors. 
However, a contribution to the overall systematic uncertainty can arise from small sector-by-sector relative 
acceptance effects. To quantify such a contribution, acceptance-corrected event yields were determined for 
each sector and compared with the average for all six sectors~\cite{CLAS-NOTE-2017-002}. The resulting 
sector-by-sector systematic uncertainty of 5.9\,\% was found to be consistent with the systematic uncertainty 
quoted for CLAS-g11a~\cite{Williams:2009yj}. The trigger efficiency maps discussed in 
Sec.~\ref{Subsection:MonteCarlo} were functions of only sector-related quantities. Therefore, the same 
uncertainty of 5.9\,\% was considered for the average trigger efficiency correction of about 18\,\%.

The contribution from the liquid hydrogen target to the overall systematic uncertainty accounts for effects 
such as the contraction, length, etc. Previous CLAS experiments have determined that the effect is approximately 
at the 0.5\,\%~level~\cite{CLAS-NOTE-2017-002}. Finally, the systematic uncertainty associated with the 
incident-photon flux normalization was estimated by studying the distribution of flux-normalized event yields 
for all production runs at different electron-beam currents~\cite{CLAS-NOTE-2017-002}.

\section{\label{Section:Results}Experimental Results}
The cross section data presented in this section have been analyzed by varying the energy-bin width in three 
different energy ranges to adjust for the available statistics in this experiment and to facilitate 
the comparison with other published data. Differential cross sections in terms of d$\sigma$/d$\Omega$ are shown
for all energies in Figs.~\ref{Figure:W-1760-1880}--\ref{Figure:W-2360-3160-forward}. Representations in terms 
of $W$ and momentum transfer $-t$ are given in Figs.~\ref{Figure:t-2520-3120-forward}--\ref{Figure:W-2520-3120-t}.
The uncertainty associated with each data point comprises contributions from the statistical uncertainty and the 
$Q$-value correlation uncertainty added in quadrature. 

\subsection{Differential cross sections d$\sigma$/d$\Omega$}
\begin{figure}[t]
 \includegraphics[width=0.5\textwidth,height=0.26\textheight]{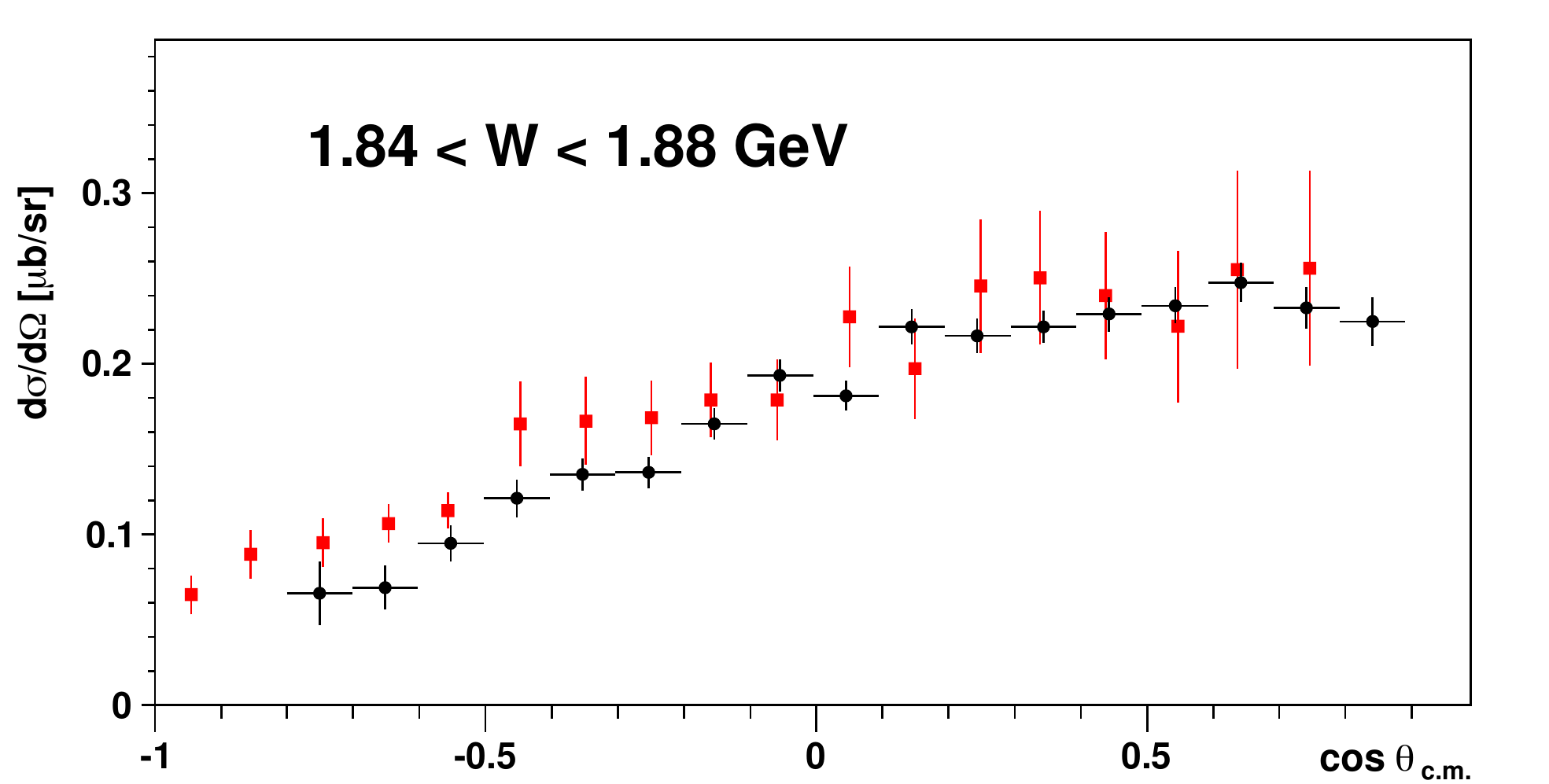}
 \caption{\label{Figure:ComparisonGRAAL} (Color online) The differential cross sections d$\sigma$/d$\Omega$
    for the energy~bin $1.84 < W < 1.88$~GeV. The new CLAS data are shown as the black solid
    circles~({\large $\bullet$}) and the uncertainties associated with each point comprise the statistical
    uncertainty and contributions from the $Q$-value correlation uncertainty added in quadrature.
    Also shown for comparison are data from the GRAAL Collaboration~\cite{Bartalini:2007fg}~({\tiny\color{red}
    {$\blacksquare$}}).}
\end{figure}

Figure~\ref{Figure:W-1760-1880} shows the differential cross sections d$\sigma$/d$\Omega$ for the $W$~range
$[\,1.76,1.88\,]$~GeV in 40-MeV-wide energy bins and 0.1-wide angle bins in cos\,$\theta_{\rm \,c.m.}^{\,\eta}$
of the $\eta$~meson in the center-of-mass frame. The CLAS-g12 data are given as the black data points. For
comparison, the distributions also show the earlier published CLAS-g11a data~\cite{Williams:2009yj} as the 
red points. These data are available in 20-MeV-wide energy bins and therefore, adjacent bins were averaged. 
The agreement is very good within the given uncertainties. Moreover, data from the A2~Collaboration at
MAMI~\cite{Kashevarov:2017kqb} are shown as the blue points with the published center-of-bin $W$~energy closest
to any of the center-of-bin energies presented in the figure. The overall agreement is good. Finally, data from
CBELSA/TAPS~\cite{Crede:2009zzb} are given as the green points. Again, the overall agreement of all four data 
sets ranges from fair to very good. Some discrepancies can be attributed to small energy mismatches in the 
presentation of the data. The CLAS-g12 data tend to be systematically lower in the backward direction for 
cos\,$\theta_{\rm\,c.m.}^{\,\eta} < -0.5$. A possible explanation is the poor CLAS acceptance in this 
kinematic range since the target was significantly shifted upstream for this experiment. The A2 and 
CBELSA/TAPS data seem to slightly underestimate the CLAS data in the forward direction for $1.80<W<1.84$~GeV. 
However, no significant normalization discrepancy is observed in any of these $W$~bins.

Figure~\ref{Figure:ComparisonGRAAL} shows a comparison of the new CLAS data (black points) with previous
results from the GRAAL Collaboration~\cite{Bartalini:2007fg} (red points) for the center-of-mass energy bin
$1.84 < W < 1.88$~GeV. The agreement is excellent in the forward direction and no overall normalization
discrepancy is observed. However, the CLAS-g12 data are again found to be slightly lower in the backward
direction for cos\,$\theta_{\rm \,c.m.}^{\,\eta} < -0.2$.

\begin{figure}[t]
 \includegraphics[width=0.5\textwidth,height=0.26\textheight]{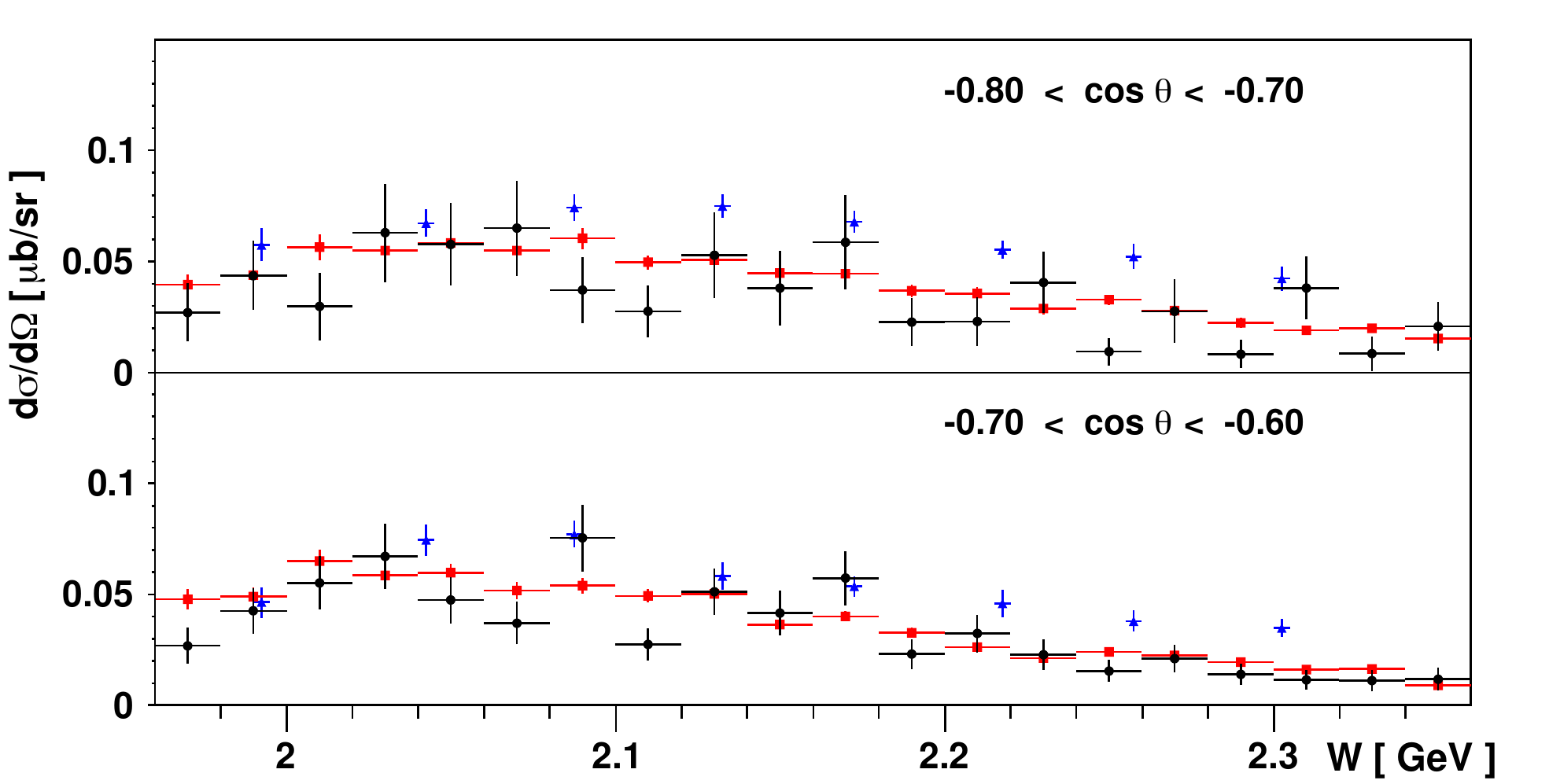}
 \caption{\label{Figure:ComparisonLEPS} (Color online) The differential cross sections d$\sigma$/d$\Omega$
    for the backward angle~bins $-0.8 < {\rm cos}\,\theta < -0.7$ (top) and $-0.7 < {\rm cos}\,\theta < -0.6$
    (bottom). The new CLAS data are shown as the black solid circles~({\large $\bullet$}) and the uncertainties
    associated with each point comprise the statistical uncertainty and contributions from the $Q$-value
    correlation uncertainty added in quadrature. Also shown for comparison are data from
    CLAS-g11a~\cite{Williams:2009yj}~({\tiny\color{red} {$\blacksquare$}}) and from the LEPS
    Collaboration~\cite{Sumihama:2009gf}~(\bl{$\blacktriangle$}).}
\end{figure}

\begin{figure*}[t]
 \includegraphics[width=1.0\textwidth]{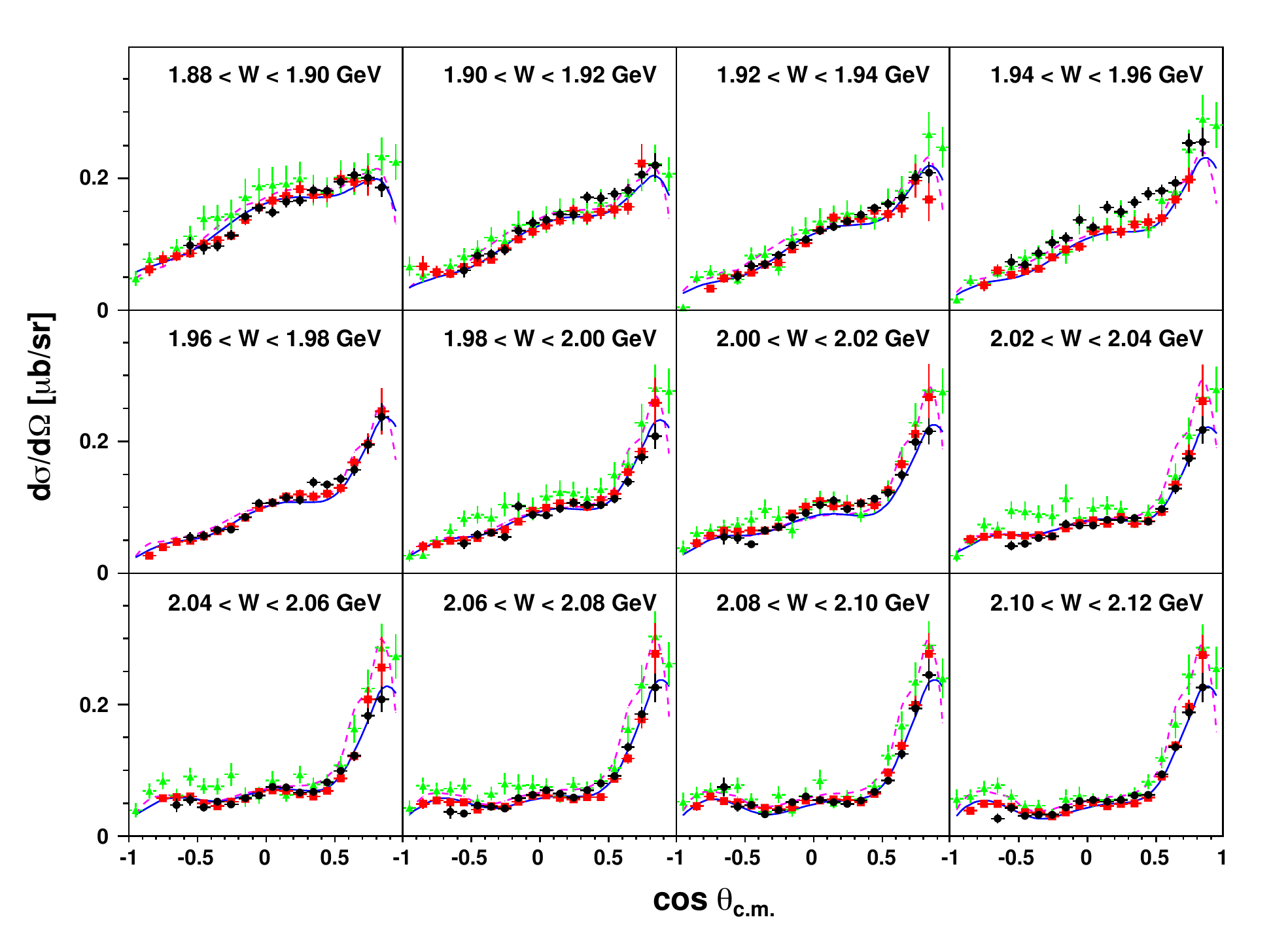}
 \caption{\label{Figure:W-1880-2120} (Color online) The differential cross sections d$\sigma$/d$\Omega$
    in 20-MeV-wide center-of-mass bins for $W\in [\,1.88,\,2.12\,]$~GeV. The new CLAS data are shown
    as the black solid circles ({\large $\bullet$}) and the uncertainties associated with each point comprise
    the statistical uncertainty and contributions from the $Q$-value correlation uncertainty added in
    quadrature. Also shown for comparison are data from CLAS-g11a~\cite{Williams:2009yj} ({\tiny\color{red}
    {$\blacksquare$}}) and from the CBELSA/TAPS Collaboration at ELSA~\cite{Crede:2009zzb}
    ({\color{green}{$\blacktriangle$}}). The blue solid and purple dashed curves denote the $\eta$-MAID\,2018
    \cite{Tiator:2018heh} and the BnGa\,2019~\cite{Muller:2019qxg} description of the
    $\gamma p\to p\eta$~cross section, respectively.}
\end{figure*}

\begin{figure*}[t]
 \includegraphics[width=1.0\textwidth]{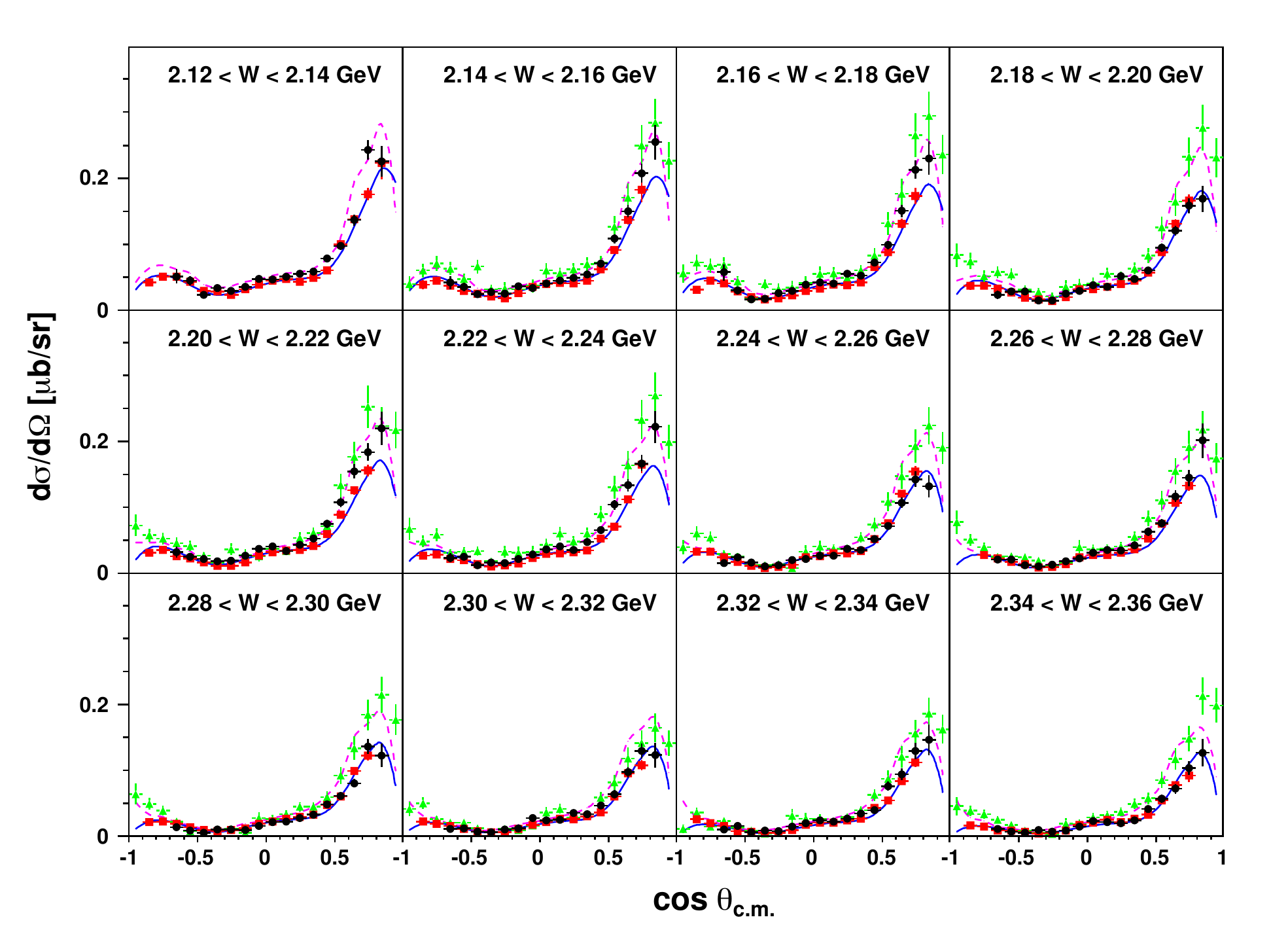}
 \caption{\label{Figure:W-2120-2360} (Color online) The differential cross sections d$\sigma$/d$\Omega$
    in 20-MeV-wide center-of-mass bins for $W\in [\,2.12,\,2.36\,]$~GeV. The new CLAS data are shown as
    the black solid circles ({\large $\bullet$}) and the uncertainties associated with each point comprise
    the statistical uncertainty and contributions from the $Q$-value correlation uncertainty added in
    quadrature. Also shown for comparison are data from CLAS-g11a~\cite{Williams:2009yj} ({\tiny\color{red}
    {$\blacksquare$}}) and from the CBELSA/TAPS Collaboration at ELSA~\cite{Crede:2009zzb}
    ({\color{green}{$\blacktriangle$}}). The blue solid and purple dashed curves denote the $\eta$-MAID\,2018
    \cite{Tiator:2018heh} and the BnGa\,2019~\cite{Muller:2019qxg} description of the
    $\gamma p\to p\eta$~cross section, respectively.}
\end{figure*}

The set of angular distributions for the energy range $W\in [\,1.88,\,2.36\,]$~GeV corresponding to the
incident photon energy $E_\gamma\in [\,1.41,\,2.50\,]$~GeV is shown in Figs.~\ref{Figure:W-1880-2120}
and~\ref{Figure:W-2120-2360} in 20-MeV-wide $W$~bins and 0.1-wide angle bins in
cos\,$\theta_{\rm \,c.m.}^{\,\eta}$. For comparison, as before, the CLAS-g11a~\cite{Williams:2009yj} and 
CBELSA/TAPS~\cite{Crede:2009zzb} data are also shown; MAMI data are only available below $E_\gamma < 1.45$~GeV 
and therefore, are not included in these figures. The earlier CLAS data have not been averaged for these 
distributions since they were published in 20-MeV-wide bins. While the agreement of the two CLAS data sets 
is excellent, the CBELSA/TAPS data tend to be systematically higher. The CBELSA/TAPS~data had to be converted 
to $W$~bins and for this reason, some discrepancies can be explained in terms of small energy mismatches. 
Nevertheless, the ELSA data seem to be systematically higher especially in the very forward and backward 
direction above $E_\gamma\approx 2.0$~GeV or $W\approx 2.2$~GeV. This observation was already discussed in 
Refs.~\cite{Williams:2009yj,Crede:2009zzb} and was also reported for other reactions, e.g., in 
$\omega$~photoproduction~\cite{Wilson:2015uoa}. The latter suggests an energy-dependent normalization issue 
of unknown nature but it is also worth emphasizing that the calorimeter-based CBELSA/TAPS experimental setup 
has better acceptance in the very forward direction. Given the excellent agreement of the two CLAS data sets, 
the reason for this discrepancy remains unclear, though.

The shapes of the angular distributions are indicative of nucleon resonance production in the entire energy
range presented in Figs.~\ref{Figure:W-1880-2120} and~\ref{Figure:W-2120-2360}. Moreover, the very prominent
forward-peaking develops around and above $W\approx 1.96$~GeV, which suggests that $t$-channel processes
become increasingly relevant.

Figure~\ref{Figure:ComparisonLEPS} shows the energy dependence of these new CLAS data (black points) for 
two backward angle bins in comparison with results from CLAS-g11a~\cite{Williams:2009yj} (red points) and from 
the LEPS Collaboration~\cite{Sumihama:2009gf} (blue points). Note that the acceptance of this CLAS experiment 
was significantly reduced in the backward direction since the liquid hydrogen target was moved upstream. The 
agreement between the two CLAS data sets within their respective uncertainties is excellent, though. The 
\mbox{SPring-8/}LEPS facility has also better acceptance for the detection of mesons in the backward direction. 
In comparison, the LEPS data points are observed to be higher than the CLAS points for the entire energy range 
shown in the figure. Moreover, the LEPS data in Fig.~\ref{Figure:ComparisonLEPS} indicate the presence of a bump 
structure around $W\approx 2.1$~GeV. In Ref.~\cite{Sumihama:2009gf}, the authors claim that a contribution from 
nucleon resonances is needed to explain this structure. Both CLAS data sets are consistent with the presence of 
this bump structure. However, the CLAS-g12 acceptance and the available statistics in this kinematic region does 
not facilitate further studies into the nature of the bump.

\begin{figure*}[t]
 \includegraphics[width=1.0\textwidth,height=0.57\textheight]{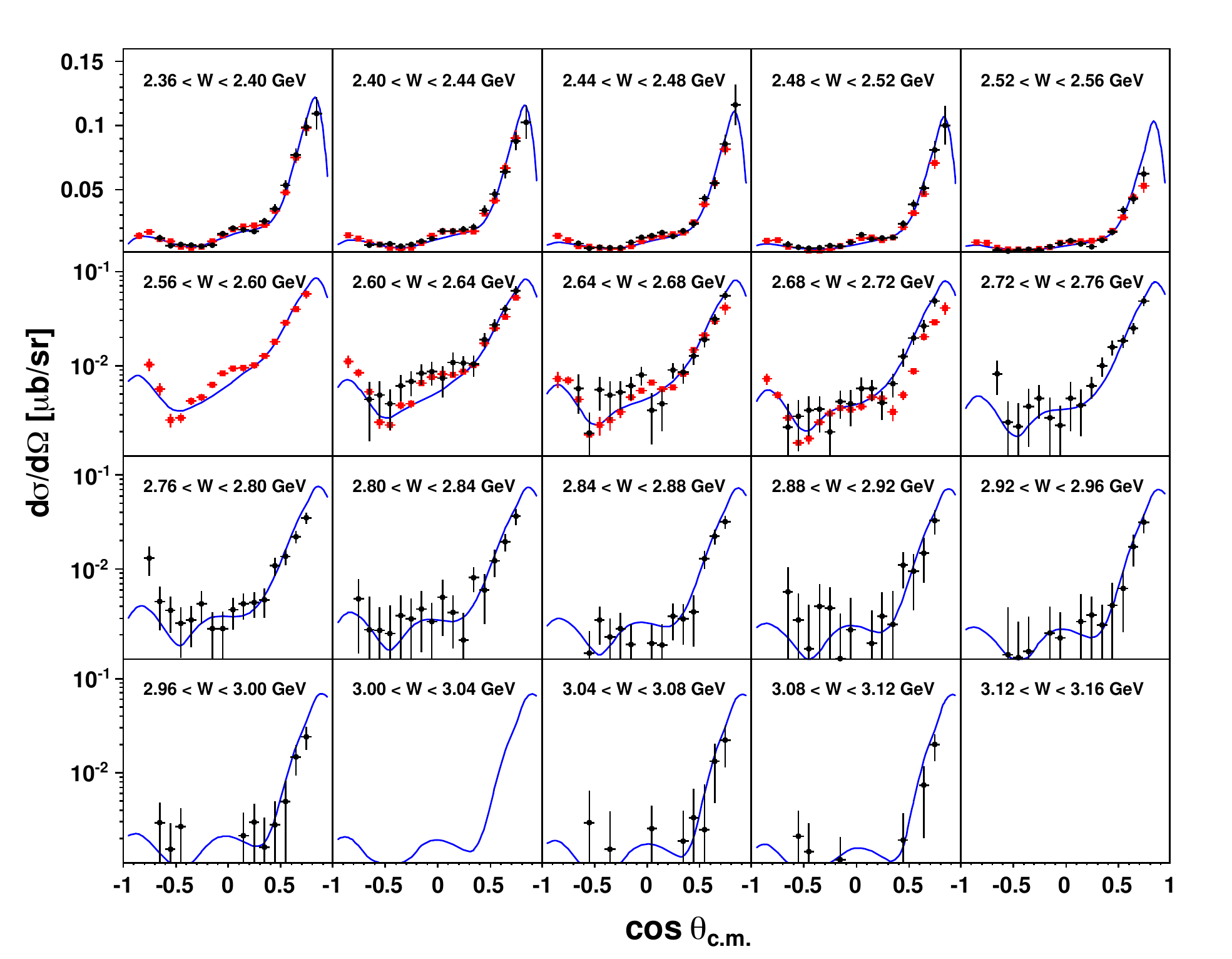}
 \caption{\label{Figure:W-2360-3160} (Color online) The differential cross sections d$\sigma$/d$\Omega$
    in 40-MeV-wide center-of-mass bins for $W\in [\,2.36,\,3.12\,]$~GeV. The new CLAS data are shown 
    as black solid circles ({\large $\bullet$}) and the uncertainties associated with each point comprise 
    the statistical uncertainty and contributions from the $Q$-value correlation uncertainty added in 
    quadrature. Also shown for comparison are data from CLAS-g11a~\cite{Williams:2009yj} ({\tiny\color{red}
    {$\blacksquare$}}). The blue solid curve denotes the $\eta$-MAID\,2018~\cite{Tiator:2018heh} description 
    of the $\gamma p\to p\eta$~cross section.}
\end{figure*}

\begin{figure*}[t]
 \includegraphics[width=1.0\textwidth,height=0.56\textheight]{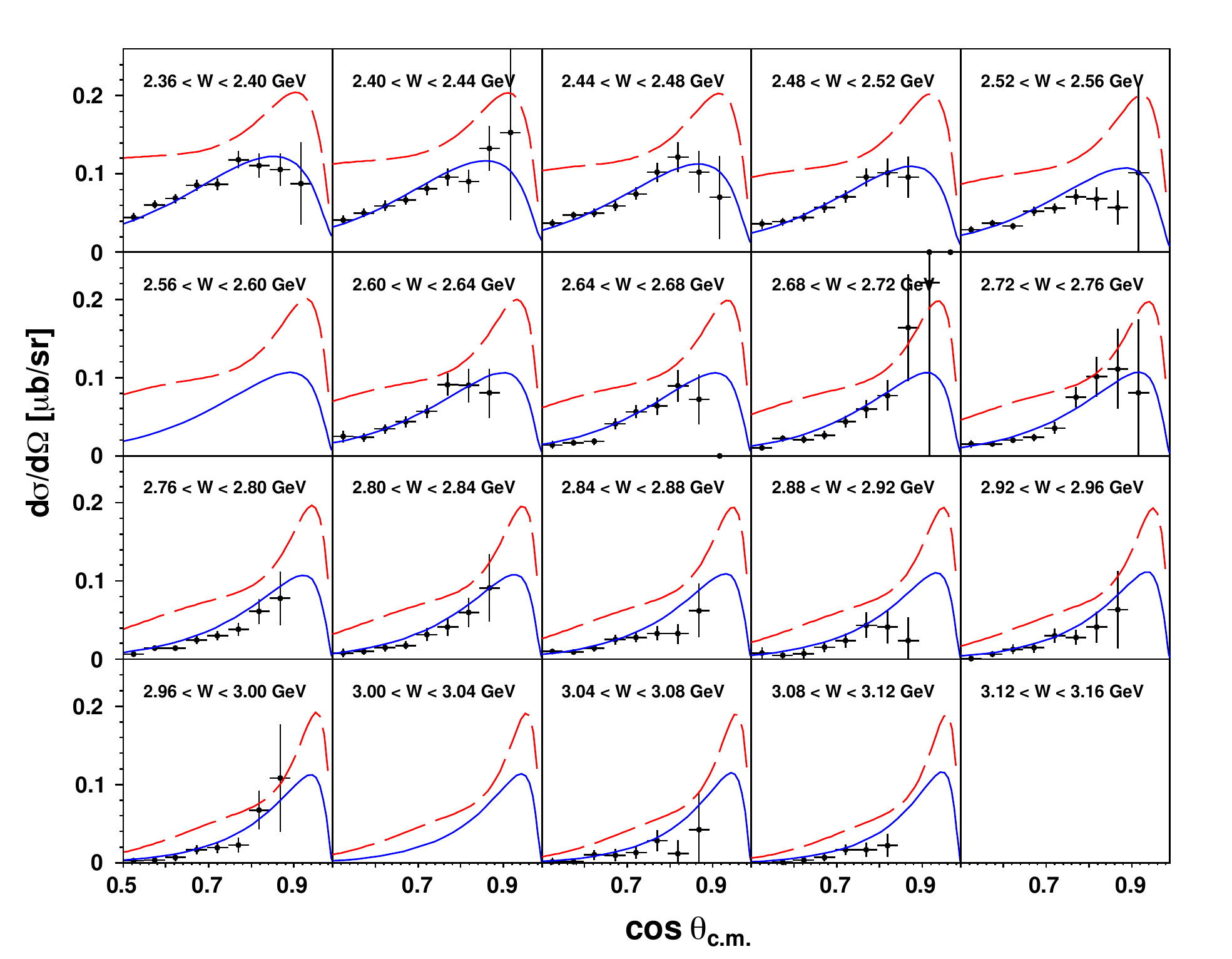}
 \caption{\label{Figure:W-2360-3160-forward} (Color online) The differential cross sections
    d$\sigma$/d$\Omega$ in 40-MeV-wide center-of-mass bins for $W\in [\,2.36,\,3.12\,]$~GeV and just the 
    forward direction cos\,$\theta_{\rm \,c.m.}^{\,\eta} > 0.5$. The new CLAS data are shown as the black 
    solid circles ({\large $\bullet$}) and the uncertainties associated with each point are comprised of 
    the statistical uncertainty and contributions from the $Q$-value correlation uncertainty added in 
    quadrature. The blue solid curve denotes the $\eta$-MAID\,2018 description~\cite{Tiator:2018heh} of 
    the $\gamma p\to p\eta$ cross section, whereas the red long-dashed curve represents the Regge model 
    discussed in Ref.~\cite{Nys:2016vjz}.}
\end{figure*}

Finally, differential cross section results for the energy range $W\in [\,2.36,\,3.12\,]$~GeV corresponding 
to incident photon energy $E_\gamma\in [\,2.50,\,4.71\,]$~GeV are shown in Fig.~\ref{Figure:W-2360-3160} in 
40-MeV-wide $W$~bins and 0.1-wide angle bins in cos\,$\theta_{\rm \,c.m.}^{\,\eta}$. Note that the vertical 
axis switches from a linear to a logarithmic scale for $W > 2.56$~GeV (second row), seemingly changing the 
shape of the angular distributions and visibly increasing the reported uncertainties. The agreement with the
CLAS-g11a data remains very good. Above $W = 2.72$~GeV ($E_\gamma \approx 3.5$~GeV), the presented data are 
first measurements. CLAS-g12 data are not available for the energy bin $2.56 < W < 2.60$~GeV, caused by 
an established tagger inefficiency in the detectors of the tagger focal plane in this region, and for the
energy bin $3.00 < W < 3.04$~GeV because the total number of events in this bin was smaller than $N_c = 500$ 
(see Sec.~\ref{Subsection:Background}) and the background subtraction (using the $Q$-factor method) could 
not be performed. Figure~\ref{Figure:W-2360-3160-forward} is similar to Fig.~\ref{Figure:W-2360-3160} but for 
the same $W$~range of $[\,2.36,3.12\,]$~GeV, only shows the forward direction $0.5 < {\rm cos}\,\theta_{\rm
\,c.m.}^{\,\eta} < 1.0$ using an angle bin size of 0.05, which is a factor of two smaller than the binning
used for the data shown in the previous figures. The reason for changing the binning in this representation
of the data is to study more closely the $t$-channel production of $\eta$~mesons beyond the baryon resonance
regime and to compare the measured angular distributions with the model described in Ref.~\cite{Nys:2016vjz}.

\subsection{Differential cross sections d$\sigma$/d$t$}
In an effort to study $\eta$~photoproduction beyond the baryon resonance regime, the differential cross sections
have been extracted also in a (W,~$-t$)~representation. This approach facilitates the comparison of the
data with Regge models that aim at describing the reaction in terms of the $t$-channel exchange of massive
quasi-particles. These new CLAS results are particularly important since they provide the missing data link in
the energy range $E_\gamma\in [\,3.5,\,4.5\,]$~GeV between the baryon resonance and the Regge-dominated regime.

\begin{figure*}[t]
  \includegraphics[width=1.0\textwidth]{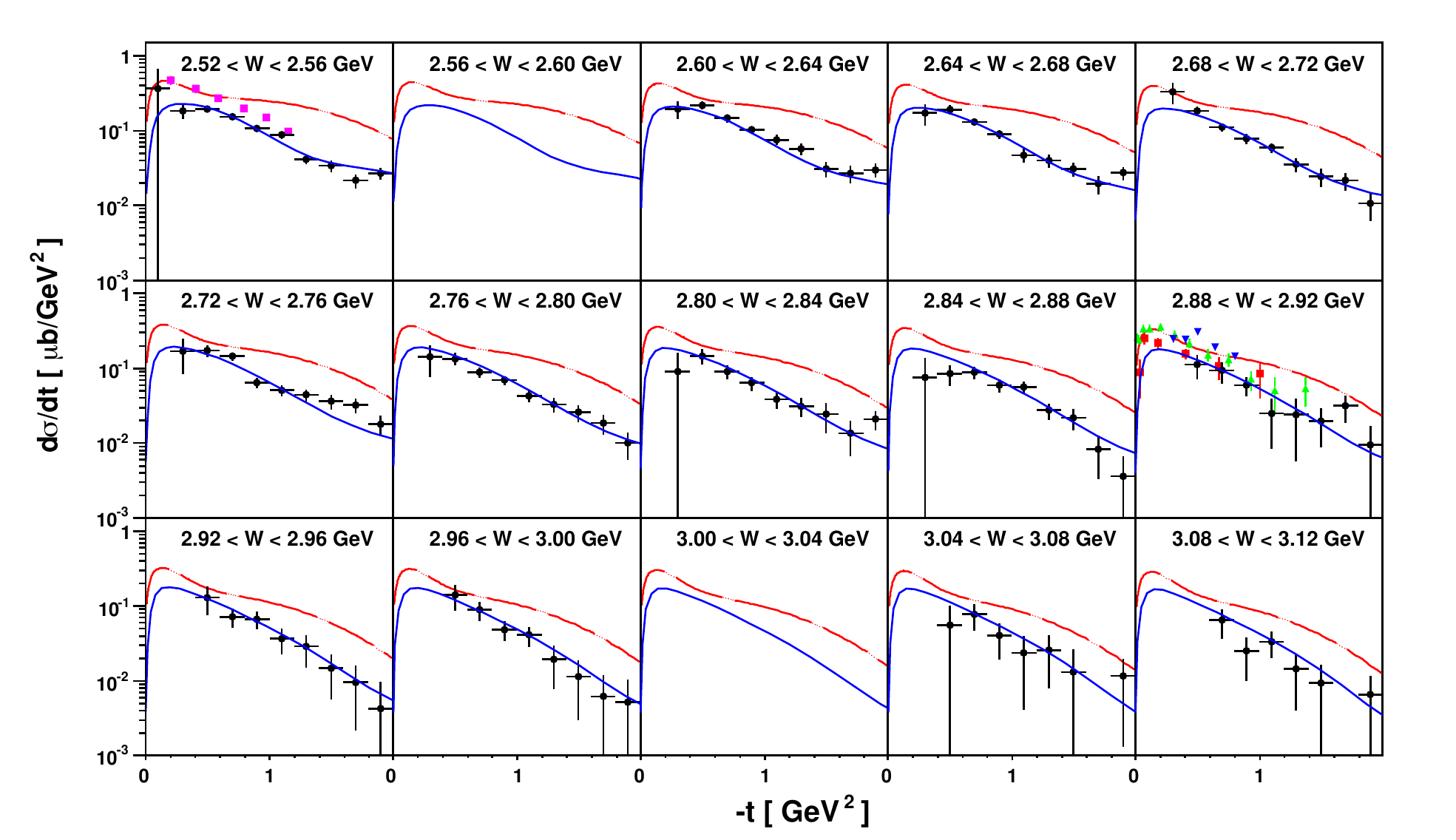}
  \caption{\label{Figure:t-2520-3120-forward} (Color online) The differential cross sections d$\sigma$/d$t$ 
    in 40-MeV-wide center-of-mass bins for $W\in [\,2.52,\,3.12\,]$~GeV and for the $-t$~range
    $[\,0,\,2\,]$~GeV$^2$. The new CLAS data are shown as the black solid circles ({\large $\bullet$}) and 
    the uncertainties associated with each point are comprised of the statistical uncertainty and contributions 
    from the $Q$-value correlation uncertainty added in quadrature. In the energy range $2.52 < W < 2.56$~GeV,
    data from NINA at the Daresbury Laboratory~\cite{Bussey:1976si}~({\tiny\color{magenta} {$\blacksquare$}})
    are given for comparison. And in the energy range $2.88 < W < 2.92$~GeV, also shown are data from 
    DESY~\cite{Braunschweig:1970jb}~({\color{green}{$\blacktriangle$}}), 
    MIT~\cite{Bellenger:1968zz}~({\tiny\color{red}{$\blacksquare$}}), and
    Cornell~\cite{Dewire:1972kk}~({\color{blue}{$\blacktriangledown$}}). The red long-dashed curve 
    represents the Regge model discussed in Ref.~\cite{Nys:2016vjz} and the blue solid curve denotes
    the $\eta$-MAID\,2018 description~\cite{Tiator:2018heh}.}
\end{figure*}

\begin{figure}[b]
 \includegraphics[width=0.49\textwidth,height=0.22\textheight]{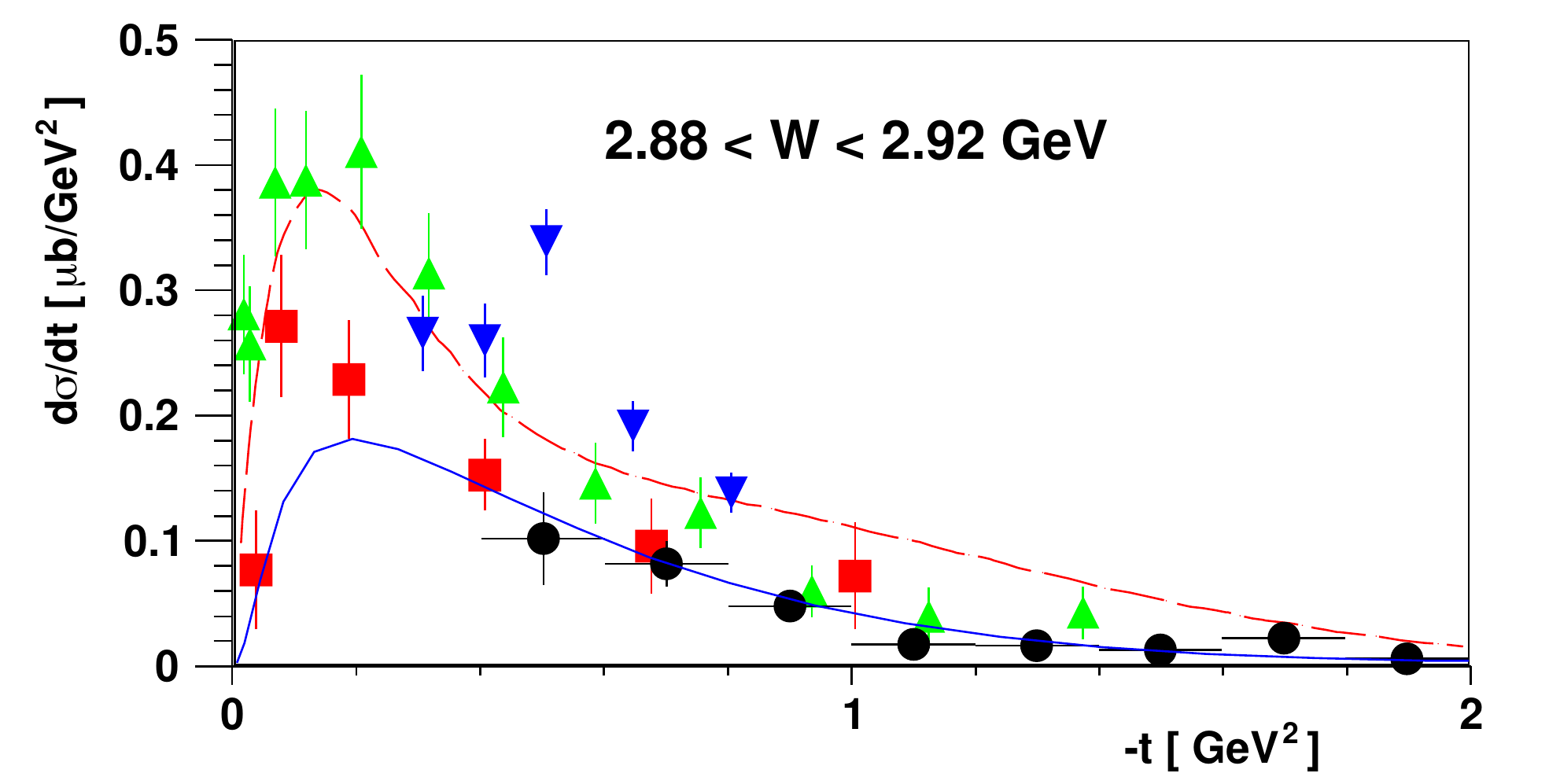}
 \caption{\label{Figure:W_2880_2920_t} (Color online) The differential cross section d$\sigma$/d$t$ for
    $W\in [\,2.88,\,2.92\,]$~GeV and for the $-t$~range $[\,0,\,2\,]$~GeV$^2$ using a linear scale. For
    the color code and an explanation of the curves, see the caption of Fig.~\ref{Figure:t-2520-3120-forward}.}
\end{figure}

Figure~\ref{Figure:t-2520-3120-forward} shows the differential cross sections d$\sigma$/d$t$ for the energy
range $W\in [\,2.52,\,3.12\,]$~GeV corresponding to incident photon energies $E_\gamma\in [\,2.91,\,4.72\,]$~GeV
using 0.2-GeV$^2$-wide $-t$~bins for $0< -t < 2$~GeV$^2$. Also shown in the figure are older data from NINA 
at the Daresbury Laboratory~\cite{Bussey:1976si} for the energy bin $2.52 < W < 2.56$~GeV as well as data from 
DESY~\cite{Braunschweig:1970jb}, the Cambridge Electron Accelerator at MIT~\cite{Bellenger:1968zz}, and 
Cornell~\cite{Dewire:1972kk}, which are only available at $W=2.9$~GeV. For the higher $W$~bin, the comparison 
between the data from the 1960s and 1970s, and the CLAS data is also presented in Fig.~\ref{Figure:W_2880_2920_t} 
using a linear scale. The MIT data are consistent with the new CLAS results, whereas all other data are found to 
be significantly higher. The older data from DESY and Cornell were used to constrain the model developed by the 
Joint Physics Analysis Center (JPAC)~\cite{Nys:2016vjz} for $-t < 1.0$~GeV$^2$ at these fairly low energies in 
the Regge regime. This model is shown as a red long-dashed curve. While the Regge model of Ref.~\cite{Nys:2016vjz} 
describes the DESY and Cornell low-$t$ data fairly well, the prediction clearly overestimates the experimental 
data points for $-t > 1.0$~GeV$^2$. The observed discrepancy between the older data (from DESY and Cornell) and 
the new CLAS data is indicative of the scale discrepancy between these new data and the JPAC curve.

\begin{figure*}[t]
 \includegraphics[width=1.0\textwidth,height=0.47\textheight]{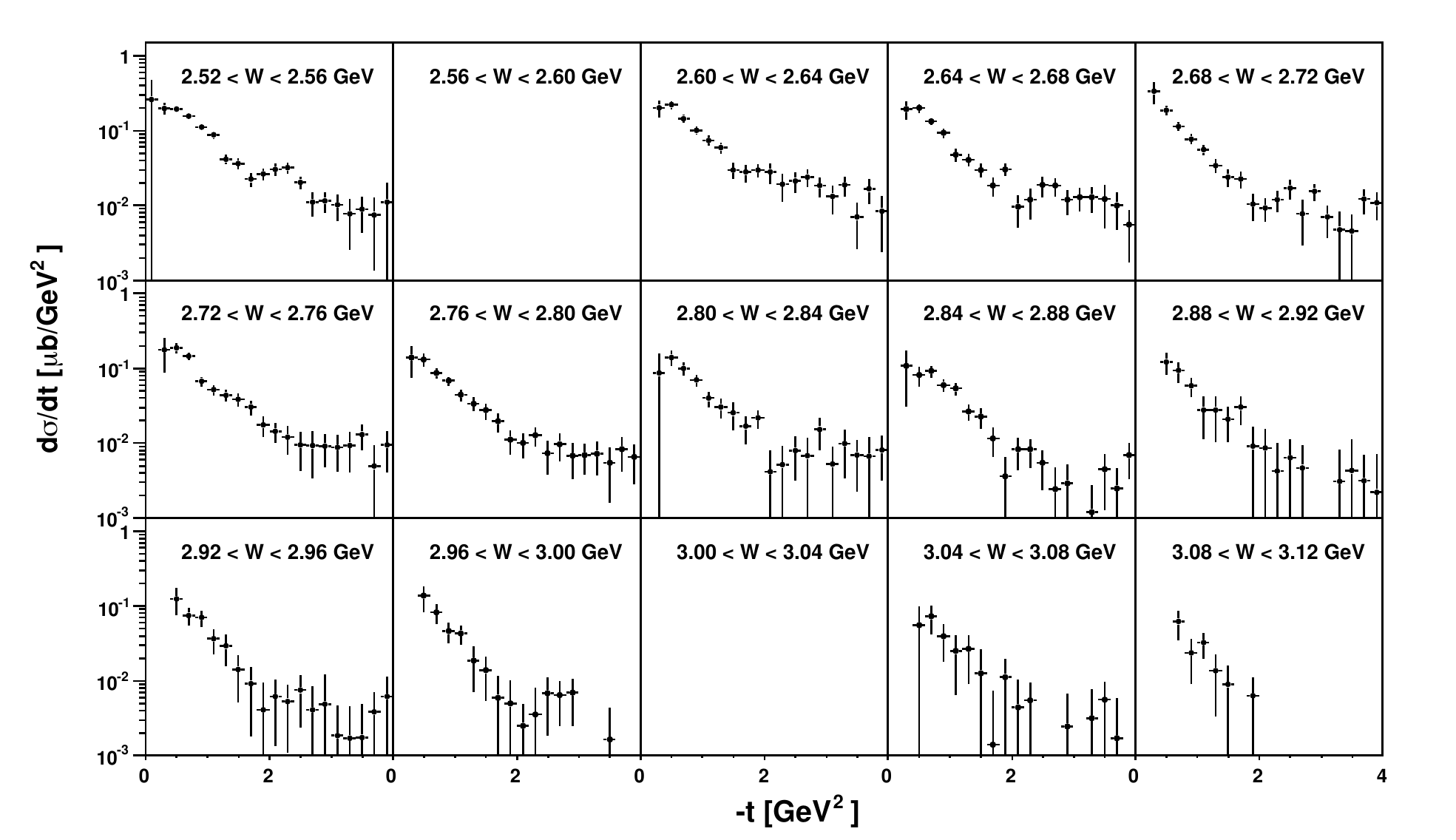}
 \caption{\label{Figure:W-2520-3120-t} The differential cross sections d$\sigma$/d$t$ in 40-MeV-wide
    center-of-mass bins for $W\in [\,2.52,\,3.12\,]$~GeV and for the full range $-t\in [\,0,\,4\,]$~GeV$^2$.
    The new CLAS data are shown as the black solid circles ({\large $\bullet$}) and the uncertainties associated
    with each point are comprised of the statistical uncertainty and contributions from the $Q$-value
    correlation uncertainty added in quadrature.}
\end{figure*}

The full set of new data points is shown in Fig.~\ref{Figure:W-2520-3120-t} for the entire analyzed $-t$~range,
$0< -t < 4$~GeV$^2$, on a logarithmic scale. The almost linear fall-off of the differential cross sections in 
the low $-t$ region is expected and can clearly be observed.

\subsection{Comparison with previous CLAS data}
\begin{figure}[t]
 \includegraphics[width=0.5\textwidth]{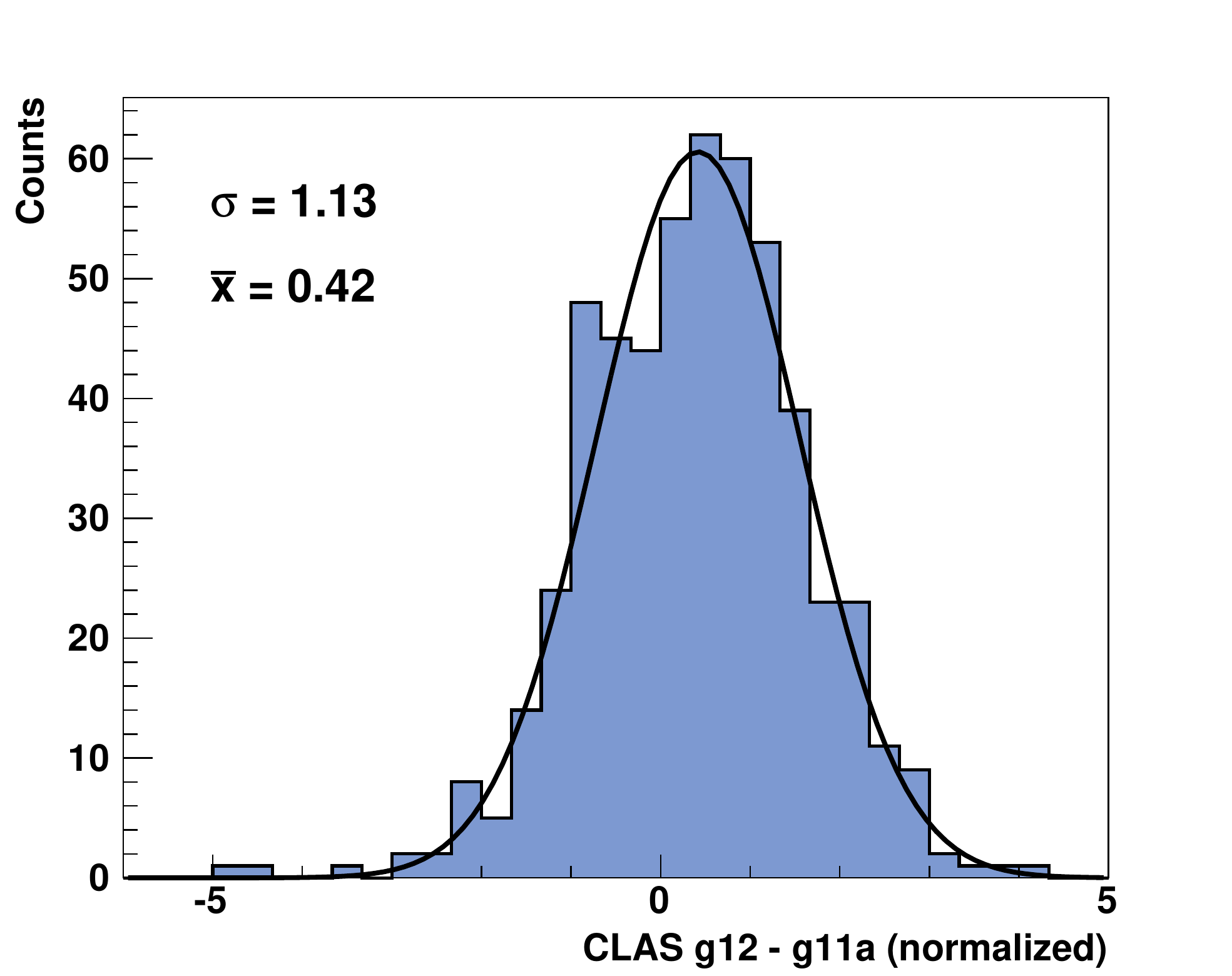}
 \caption{\label{Figure:Comparison-pull} (Color online) Comparison between the new and the
   published~\cite{Williams:2009yj} CLAS data in form of a difference distribution normalized to their
   uncertainties. See text for more details.}
\end{figure}

Figure~\ref{Figure:Comparison-pull} shows a comparison of the new CLAS data with the previously published
CLAS data on $\eta$~photoproduction~\cite{Williams:2009yj} in the form of a normalized difference distribution:
\begin{eqnarray}
  \label{Eq:pullDistribution}
  \frac{\bigl(\frac{d\sigma}{d\Omega}\bigr)_{\rm g12}\,-\,\bigl(\frac{d\sigma}{d\Omega}\bigr)_{\rm g11a}}
  {\sqrt{(\Delta\sigma)^2_{\rm g12}\,+\,(\Delta\sigma)^2_{\rm g11a}}}~,
\end{eqnarray}
where the uncertainties in the denominator are comprised only of statistical and $Q$-value correlation 
uncertainties.

With the exception of a small structure around $-1.0$ due to a possible poorly understood acceptance effect 
in either experiment, the distribution is symmetric and Gaussian indicating that any discrepancies in the 
shape of the differential cross sections are mostly statistical in nature. The Gaussian width of $\sigma = 1.13$ 
suggests that the uncertainties in the denominator of Eq.~(\ref{Eq:pullDistribution}) are slightly underestimated.

\begin{figure}[t]
 \includegraphics[width=0.5\textwidth]{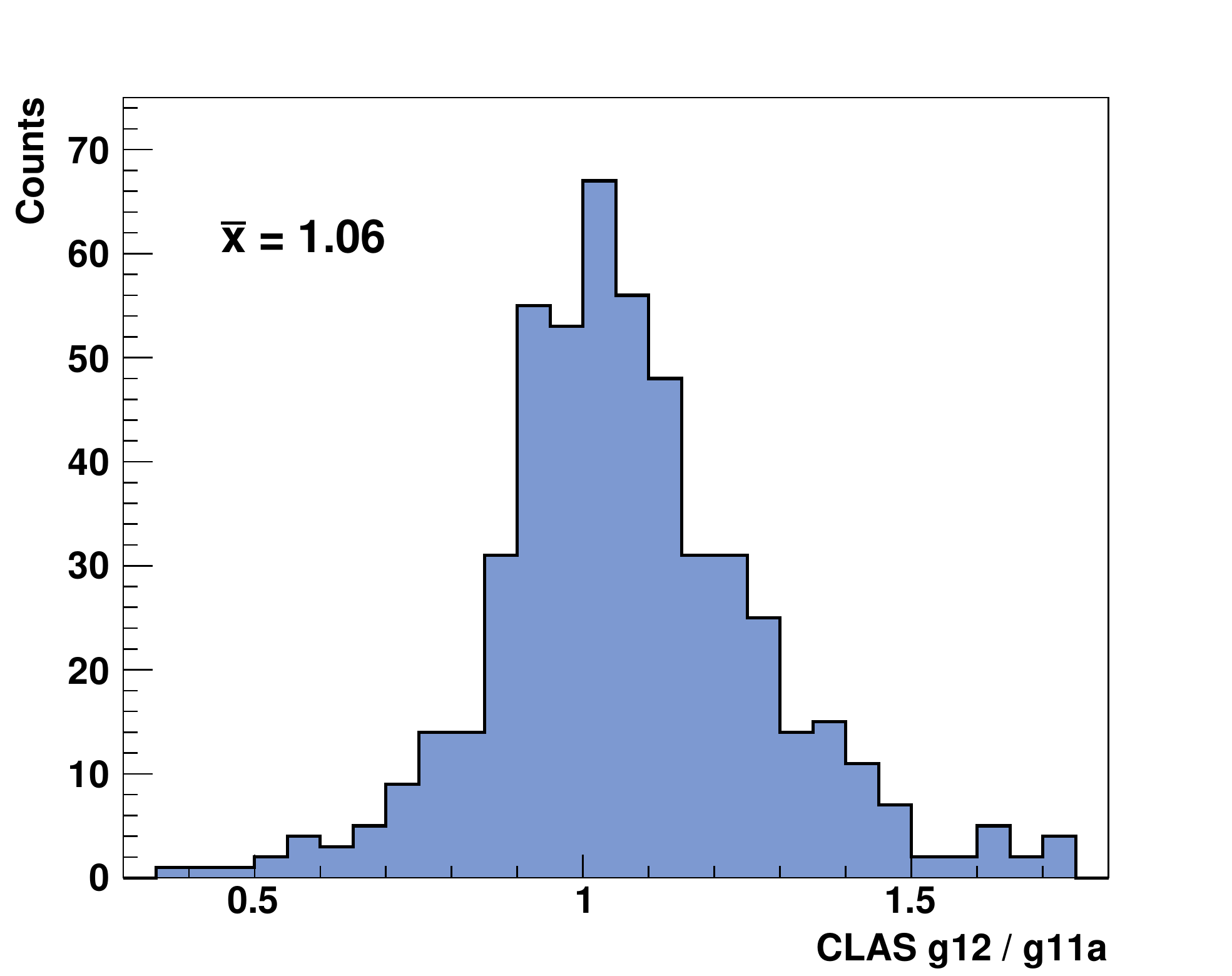}
 \caption{\label{Figure:Comparison-ratio} (Color online) Unweighted ratio distribution of
   the new and the published~\cite{Williams:2009yj} CLAS data.}
\end{figure}

As a matter of fact, no additional uncertainties are included beyond those listed in
Table~\ref{Table:Uncertainties} to guarantee consistency between the two data sets. However, the difference
distribution is slightly shifted toward positive values. Figure~\ref{Figure:Comparison-ratio} shows the
unweighted ratio distribution of the same two data sets. This distribution is also fairly symmetric with a mean 
value of 1.06, which indicates that an overall increase of about 6\,\% is observed in the new data. The electron 
rates detected by the tagger and used to compute the number of photons incident on the target are integrated over 
the live time of the experiment. In the previous CLAS measurement~\cite{Williams:2009yj}, the clock-based
live time calculation was checked by using the counts of a Faraday cup located downstream of CLAS. Despite
high statistical uncertainties in these secondary measurements, a current-dependent live time was observed and
at maximum electron beam current, the dead time was determined to be about a factor of two higher than the one
given by the clock-based measurement used for the flux normalization. The corresponding correction resulted in
the largest single-source contribution to the overall systematic uncertainty. Such a current-dependent live time
was not observed for the CLAS-g12 data. The reason for this effect in the previous CLAS experiment remains poorly 
understood. However, the observed overall scale discrepancy between the two CLAS measurements of the 
$\gamma p\to p\eta$ cross sections is well within the reported uncertainties for these two experiments.

\section{\label{Section:PWA}Physics Discussion}

Various theoretical and phenomenological approaches have been applied and studied to describe
$\eta$~photoproduction on the nucleon, in particular to understand nucleon resonance contributions to this
reaction, e.g., effective field theory~\cite{Ruic:2011wf}, dispersion theoretical 
calculations~\cite{Aznauryan:2003zg}, and Regge models~\cite{Sibirtsev:2010yj,Kashevarov:2017vyl}. 

A special group of models are isobar models, e.g., Refs.~\cite{Chiang:2001as,Chiang:2002vq,Tryasuchev:2003st},
which treat nucleon resonances in terms of $s$-channel Breit-Wigner parametrizations using energy-dependent
widths due to their couplings with other decay channels. The non-resonant background amplitude is typically
written as a sum of Born terms and $t$-channel meson-exchange contributions. In $\eta$~photoproduction,
Born terms are usually suppressed because their coupling constants are fairly small. In such isobar models,
the double-counting of terms due to the quark-hadron duality is often concerning since the sum of an infinite
series of $s$-channel resonances is equivalent to an infinite sum of $t$-channel meson-exchange amplitudes.
In the $\eta$-MAID\,2018 isobar model described in Ref~\cite{Tiator:2018heh}, the double-counting is removed
by introducing a damping factor to the Regge amplitudes. Moreover, despite the minor role of Born terms, their
couplings are determined from fitting experimental data.

The latest $\eta$-MAID\,2018 solution is shown as a blue solid curve in
Figs.~\ref{Figure:W-1760-1880}--\ref{Figure:W_2880_2920_t}. The experimental data are described very well
over the entire energy range. All known $N^\ast$~states listed in the RPP~\cite{Tanabashi:2018oca} were used 
to describe the resonance regime from the $\gamma p\to p\eta$~threshold up to $W < 2.5$~GeV. For a given
partial wave $\alpha$, the set of $N_\alpha$ nucleon resonances were added as generalized Breit-Wigner functions
with a unitary phase $\phi$ for each resonance~\cite{Tiator:2018heh}. For the higher-energy regime, Regge 
phenomenology was applied and in an effort to provide a continuous description of $\eta$~photoproduction from 
threshold to about $W\approx 5$~GeV, the damping fator was introduced, which goes to zero at the $\eta$~production
threshold and approaches unity above $W = 2.5$~GeV, where Regge description fully sets in~\cite{Tiator:2018heh}.
Only two resonances were found to be insignificant in their contribution to $\gamma p\to p\eta$: The 
$N(2040)\,3/2^+$ resonance, a one-star state observed by BES\,II in $J/\psi$~decays to 
$N\overline{N}\pi$~\cite{Ablikim:2004ug,Ablikim:2009iw}, and the $N(2220)\,9/2^+$~resonance. In their description,
the reaction is dominated by the $1/2^-$~partial wave that is associated with contributions from the 
$N(1535)\,1/2^-$, $N(1650)\,1/2^-$, and $N(1895)\,1/2^-$ states. In the fourth resonance region, the most 
significant contributions beyond the $1/2^-$~partial wave come from the $N(1875)\,3/2^-$, $N(1900)\,3/2^+$, 
and the $N(1860)\,5/2^+$~nucleon resonances.

The multi-channel Bonn-Gatchina (BnGa) partial wave analysis (PWA) uses a large experimental database, which
includes data on pion- and photo-induced meson-production reactions, with up to two pseudoscalar mesons in
the final state~\cite{BnGa:Database}. The approach is based on a fully relativistically invariant operator 
expansion method and combines the analysis of different reactions imposing directly analyticity and unitarity 
constraints~\cite{Beck:2016hcy}. Figures~~\ref{Figure:W-1760-1880}--\ref{Figure:W-2120-2360} show the BnGa 
solution BnGa\,2019 as a purple curve; more details are discussed in Ref.~\cite{Beck:2016hcy}. Overall, the 
BnGa curve describes the experimental data very well. Deviations from the $\eta$-MAID\,2018 solution can be 
observed, mostly in the forward direction above $W\approx 2$~GeV. The difference between the two curves can 
be traced back to very similar discrepancies between the CLAS and CBELSA/TAPS data sets, which received 
different weights in the interpretation of the data. The $1/2^-$~partial wave also dominates the BnGa 
description of the $\gamma p\to p\eta$ reaction. However, in the fourth resonance region, the 
$N(1900)\,3/2^+$ resonance plays a significantly more important role than in $\eta$-MAID\,2018, whereas 
contributions from the other two states found significant in $\eta$-MAID, $N(1875)\,3/2^-$ and 
$N(1860)\,5/2^+$, are practically negligible~\cite{Anisovich:2017afs}. The identification of significant 
contributions from different nucleon resonances in $\eta$~photoproduction is not surprising since the 
polarization observables are still scarce.

The high-energy regime above $E_\gamma = 4$~GeV is studied in terms of the Regge amplitudes discussed in 
Ref.~\cite{Nys:2016vjz}. While each Regge exchange has a known energy dependence, the $t$~behavior is 
{\it a priori} unknown. In the approach of the Joint Physics Analysis Center (JPAC)~\cite{Nys:2016vjz}, 
information from the resonance region is used through dispersion relations and finite-energy sum rules (FESR) 
to extract the $t$-dependence of the differential cross sections at high energies. Specifically, the 
$\eta$-MAID\,2001 model~\cite{Chiang:2001as} was used for the low-energy parametrizations. Data from 
DESY~\cite{Braunschweig:1970jb} and Cornell~\cite{Dewire:1972kk} for $0< -t < 1$~GeV$^2$ were used to 
constrain the JPAC model. The available data sets and the model are shown in Fig.~\ref{Figure:W_2880_2920_t} 
for $E_\gamma \approx 4.0$~GeV. The red JPAC curve describes the DESY and the Cornell data well but is observed 
to be systematically off in its description of the new CLAS data and the data from MIT~\cite{Bellenger:1968zz}. 
This scaling problem is also observed in Fig.~\ref{Figure:t-2520-3120-forward}.

For the Regge exchanges in the JPAC approach, two high-energy models have been developed. In the first more
conservative approach, shown in Figs.~\ref{Figure:W-2360-3160-forward}--\ref{Figure:W_2880_2920_t}, only 
$t$-channel exchanges associated with the known ($J^{PC} = 1^{--}$) vector- ($\omega,\,\rho$) and ($J^{PC} = 
1^{+-}$) axial-vector ($h_1,\,b_1$) meson resonances are considered, whereas the second model includes exchanges 
that correspond to, as yet, unobserved mesons. The latter approach explores the possible impact of a 
$2^{--}$~exchange that would result in increased cross sections and in a beam asymmetry smaller than one. The 
conservative JPAC predictions are observed to systematically overestimate the new CLAS data and the inclusion 
of a $2^{--}$~exchange would only increase this discrepancy. The preliminary conclusion is that these new data 
and more importantly, recent results on the beam asymmetry in $\eta$~photoproduction at high energies reported 
by the GlueX Collaboration~\cite{AlGhoul:2017nbp,Adhikari:2019gfa} are in clear contradiction with these 
predictions. The $2^{--}$~exchange has also not been considered in the $\eta$-MAID\,2018 model. The $C$-parity 
conservation prohibits exchanges of scalar and pseudoscalar mesons.

In conclusion, these experimental data confirm the expectation that the $\gamma p\to p\eta$ reaction proceeds 
primarily through $\rho$ and $\omega$ vector-meson $1^{--}$~exchange. Other $\eta$~data also confirm the predicted 
rapid decline of the cross sections in the very forward direction of the $\eta$~meson in the center-of-mass frame, 
which corresponds to the differential cross sections at very small values of $-t$. This behavior is clearly visible
in the model descriptions shown in Figs.~\ref{Figure:W-2360-3160-forward}--\ref{Figure:W_2880_2920_t}. For 
cos\,$\theta_{\rm \,c.m.} = 1$ or $t^{\,\prime} = t - t_{\rm \,min} = 0$~GeV$^2$, conservation of angular 
momentum requires conservation of helicities:
\begin{eqnarray}
\lambda_\gamma\,-\,\lambda_{\rm \,proton}\,=\,\lambda_{\,\eta}\,-\,\lambda_{\rm \,proton^{\,\prime}}\,,
\end{eqnarray}
where the right-hand side denotes the helicity of the recoiling proton with $\lambda_{\,\eta}=0$. In Regge 
models, this imposes an even stronger constraint since conservation of angular momentum is required at the 
top ($\gamma$\,-\,$\eta$) vertex and at the bottom ($N$\,-\,$N$) vertex in the right diagram of 
Fig.~\ref{Figure:feynman}. Since the helicity of a real photon, $\lambda = \pm 1$, cannot turn into 
$\lambda = 0$ for the $\eta$~meson, the amplitude needs to vanish and the cross section decreases to {\it zero}. 
In Regge pole theory, this behavior is thus built into the top vertex by factorization~\cite{Irving:1977ea}. In 
contrast, using virtual photons, the cross section in the very forward direction proceeds primarily via the 
photon's longitudinal component.

\section{\label{Section:Summary}Summary and Outlook}
Photoproduction cross sections have been presented for the reaction $\gamma p\to p\eta$ using tagged photons 
and the CLAS spectrometer at Jefferson Laboratory. The results are shown for incident photon energies between 
about 1.2 and 4.7~GeV. These new $\eta$~photoproduction data are consistent with earlier CLAS results but 
extend the energy range beyond the nucleon resonance regime. Cross sections $d\sigma/dt$ are also presented 
for $W > 2.52$~GeV and studied in terms of the dominant Regge exchange amplitudes. While axial vector exchanges 
are negligible, the data confirm the expected dominance of vector-meson exchanges. Calculations using 
finite-energy sum rules (FESR) indicate that the $2^{--}$~exchange could be relevant but predictions are 
inconsistent with the differential cross section data presented here and with beam-asymmetry results recently 
reported by the GlueX Collaboration. In light of these new CLAS-g12 data and the new \mbox{$\eta$-MAID\,2018} 
model, it would certainly be interesting to revisit the JPAC approach. Upcoming data from the GlueX Collaboration 
will extend the differential cross section measurements to $W\approx 4.2$~GeV and further prepare the foundation
for a global analysis of low- and high-energy data of related reactions within the framework of FESR.

In the baryon resonance regime, a comparison of the differential cross sections d$\sigma$/d$\Omega$ with 
predictions of the isobar model \mbox{$\eta$-MAID\,2018} and the BnGa coupled-channel analysis confirms the 
dominance of the $1/2^-$~partial wave close to the reaction threshold. The unambiguous identification of resonance 
contributions in the fourth resonance region is still challenging owing to the lack of polarization observables 
around $W\approx 2$~GeV.

\begin{acknowledgments}
% put your acknowledgments here.
The authors thank the technical staff at Jefferson Laboratory and at all the participating institutions 
for their invaluable contributions to the success of the experiment. This research is based on work
supported by the U. S. Department of Energy, Office of Science, Office of Nuclear Physics, under Contract 
No. DE-AC05-06OR23177. The group at Florida State University acknowledges additional support from the U.S.
Department of Energy, Office of Science, Office of Nuclear Physics, under Contract No. DE-FG02-92ER40735.
This work was also supported by the US National Science Foundation, the State Committee of Science of 
Republic of Armenia, the Chilean Comisi\'{o}n Nacional de Investigaci\'{o}n Cient\'{i}fica y Tecnol\'{o}gica 
(CONICYT), the Italian Istituto Nazionale di Fisica Nucleare, the French Centre National de la Recherche 
Scientifique, the French Commissariat a l'Energie Atomique, the Scottish Universities Physics Alliance 
(SUPA), the United Kingdom's Science and Technology Facilities Council, and the National Research 
Foundation of Korea. %, the Deutsche 
%Forschungsgemeinschaft (SFB/TR110), and the Russian Science Foundation
%under Grant No. 16-12-10267. 
\end{acknowledgments}

%\begin{figure}[t]
% %\includegraphics[width=0.41\textwidth,height=0.25\textheight]{FIGURES/ratio.pdf}
% \includegraphics[width=0.45\textwidth]{FIGURES/ratio.pdf}
% \caption{\label{Figure:Comparison-ratio} (Color online) Unweighted ratio distribution of
%   the new and the published~\cite{Williams:2009yj} CLAS data.}
%\end{figure}

% Create the reference section using BibTeX:
%\bibliography{basename of .bib file}

\end{document}